\documentclass[fleqn,usenatbib]{mnras}

\usepackage[T1]{fontenc}

\DeclareRobustCommand{\VAN}[3]{#2}
\let\VANthebibliography\thebibliography
\def\thebibliography{\DeclareRobustCommand{\VAN}[3]{##3}\VANthebibliography}

\usepackage{threeparttable}

\usepackage{graphicx}	
\usepackage{amsmath}	
\usepackage{amssymb}	

\usepackage{times,txfonts} 

\newcommand{\sBV}{$s_{\mathrm{BV}}$}
\newcommand{\dm}{dm$_{2}(i)$}
\newcommand{\tdm}{tdm$_{2}(i)$}
\newcommand{\magdsq}{mag\,d$^{-2}$}
\newcommand{\kms}{km~s$^{-1}$}

\title[SN I\MakeLowercase{a} $i$-band light-curve features]{Carnegie Supernova Project: kinky $i$-band light-curves of Type Ia supernovae}

\author[P. J. Pessi et al.]{
P. J. Pessi,$^{1,2}$\thanks{E-mail: pjpessi@fcaglp.unlp.edu.ar}
E. Y. Hsiao,$^{3}$
G. Folatelli,$^{1,4,5}$ 
J. P. Anderson,$^{2}$
C. R. Burns,$^{6}$
S. Uddin,$^{7}$
L. Galbany,$^{8,9}$
\newauthor
M. M. Phiilips,$^{10}$
N. Morrell,$^{10}$
C. Ashall,$^{11}$
E. Baron,$^{12}$
C. Contreras,$^{10}$
M. Hamuy,$^{13,14}$
P. Hoeflich,$^{3}$
\newauthor
K. Krisciunas,$^{7}$
S. Kumar,$^{3}$
J. Lu,$^{3}$
L. Martinez,$^{1,4,15}$
A. L. Piro,$^{6}$
M. Shahbandeh,$^{3}$
M. D. Stritzinger,$^{16}$
\newauthor
and N. B. Suntzeff$^{7}$
\\
$^{1}$Facultad de Ciencias Astron\'omicas y Geof\'isicas, Universidad Nacional de La Plata, Paseo del Bosque S/N, B1900FWA, La Plata, Argentina\\
$^{2}$European Southern Observatory, Alonso de C\'ordova 3107, Casilla 19, Santiago, Chile\\
$^{3}$Department of Physics, Florida State University, 77 Chieftan Way, Tallahassee, FL 32306, USA\\
$^{4}$Instituto de Astrof\'isica de La Plata (IALP), CCT-CONICET-UNLP. Paseo del Bosque S/N, B1900FWA, La Plata, Argentina\\
$^{5}$Kavli Institute for the Physics and Mathematics of the Universe (WPI), The University of Tokyo, 5-1-5 Kashiwanoha, Kashiwa,Chiba 277-8583, Japan\\
$^{6}$Observatories of the Carnegie Institution for Science, 813 Santa Barbara St, Pasadena, CA 91101, USA\\
$^{7}$George P. and Cynthia Woods Mitchell Institute for Fundamental Physics and Astronomy, Department of Physics and Astronomy,\\ Texas A\&M University, College Station, TX, 77843, USA\\
$^{8}$Institute of Space Sciences (ICE, CSIC), Campus UAB, Carrer de Can Magrans, s/n, E-08193 Barcelona, Spain\\
$^{9}$Institut d'Estudis Espacials de Catalunya (IEEC), E-08034 Barcelona, Spain\\
$^{10}$Carnegie Observatories, Las Campanas Observatory, Colina El Pino, Casilla 601, Chile\\
$^{11}$Institute for Astronomy, University of Hawaii, 2680 Woodlawn Drive, Honolulu, HI 96822, USA\\
$^{12}$Homer L. Dodge Department of Physics and Astronomy, 440 West Brooks Street, Room 100, Norman, OK 73019, USA\\
$^{13}$Vice President and Head of Mission of AURA-O in Chile, Avda Presidente Riesco 5335 Suite 507, Santiago, Chile\\
$^{14}$Hagler Institute for Advanced Studies, Texas A\&M University, College Station, TX 77843, USA\\
$^{15}$Universidad Nacional de R\'io Negro. Sede Andina, Mitre 630 (8400) Bariloche, Argentina\\
$^{16}$Department of Physics and Astronomy, Aarhus University, Ny Munkegade, DK-8000 Aarhus C, Denmark
}

\date{Accepted 2021 December 04. Received 2021 December 03; in original form 2021 October 28}

\pubyear{2021}

\begin{document}
\label{firstpage}
\pagerange{\pageref{firstpage}--\pageref{lastpage}}
\maketitle

\begin{abstract}
We present detailed investigation of a specific $i$-band light-curve feature in Type~Ia supernovae (SNe~Ia) using the rapid cadence and high signal-to-noise ratio light-curves obtained by the Carnegie Supernova Project.
The feature is present in most SNe~Ia and emerges a few days after the $i$-band maximum.
It is an abrupt change in curvature in the light-curve over a few days and appears as a flattening in mild cases and a strong downward concave shape, or a ``kink'', in the most extreme cases.
We computed the second derivatives of Gaussian Process interpolations to study 54 rapid-cadence light-curves.
From the second derivatives we measure: 1) the timing of the feature in days relative to $i$-band maximum; \tdm\ and 2) the strength and direction of the concavity in \magdsq\ ; \dm.
76\% of the SNe~Ia show a negative \dm, representing a downward concavity - either a mild flattening or a strong ``kink.''
The \tdm\ parameter is shown to correlate with the color-stretch parameter \sBV, a SN~Ia primary parameter.
The \dm\ parameter shows no correlation with \sBV\ and therefore provides independent information.
It is also largely independent of the spectroscopic and environmental properties.
Dividing the sample based on the strength of the light-curve feature as measured by \dm, SNe~Ia with strong features have a Hubble diagram dispersion of 0.107~mag, 0.075~mag smaller than the group with weak features. 
Although larger samples should be obtained to test this result, it potentially offers a new method for improving SN~Ia distance determinations without shifting to more costly near-infrared or spectroscopic observations.
\end{abstract}

\begin{keywords}
transients: supernovae -- methods: observational
\end{keywords}



\section{Introduction}
\label{sec:intro}

Type Ia supernovae (SNe~Ia) are key cosmic distance indicators.
The discovery of the accelerated expansion, or dark energy, was based on SN~Ia distance determination to distant galaxies \citep{1998AJ....116.1009R, 1999ApJ...517..565P}.
They remain a vital tool for characterizing the properties of dark energy.

The vast majority of SNe~Ia are remarkably uniform. 
In the optical, they are ``standardizable'' candles as their peak luminosities are found to correlate with the light-curve shape and colors \citep[e.g.,][]{1977SvA....21..675P,1996AJ....112.2438H,1998A&A...331..815T}. 
Parameterizing the light-curve shape, decline rate parameters such as $\Delta m_{15} (B)$ measured in the $B$-band have been shown to be the primary parameter of normal SNe~Ia, correlating with their peak luminosities \citep[in the sense that fainter SNe Ia have faster declining light-curves;][]{1993ApJ...413L.105P}, spectra \citep{1995ApJ...455L.147N}, and even host galaxy environments \citep[e.g.,][]{2000AJ....120.1479H, 2009ApJ...691..661H}.
The updated light-curve parameterization, color-stretch \sBV, which is measured as the time difference between the occurrence of $B$-band maximum and the reddest point in the $(B-V)$ color curve divided by 30~d \citep{2014ApJ...789...32B}, was shown to improve the standardization in the faint end but still works similarly to $\Delta m_{15} (B)$ as a primary parameter.

These empirical relations reduce the scatter on the Hubble diagram to roughly $0.15-0.20$ mag.
Considerable efforts have been focused on further improvement.
They include shifting the standardization from optical light-curves to the near-infrared (NIR) \citep[e.g.,][]{2004ApJ...602L..81K,2012MNRAS.425.1007B, 2018A&A...615A..45S} and through purely spectroscopic means \citep[e.g.,][]{2009A&A...500L..17B, 2015ApJ...815...58F}.
Additional parameters, independent of the light curve decline rate, may be found by comparing observations with models \citep[e.g.,][]{2018A&A...611A..58G} or through purely statistical means \citep[e.g.,][]{2013ApJ...766...84K}.
Nevertheless, independent secondary or tertiary parameters for SNe~Ia have remained largely elusive. 

One place to search for such an independent parameter may be in the rest-frame $i$-band region.
The region is dominated by the strong \ion{Ca}{II} infrared triplet feature \citep[e.g.,][]{2005ApJ...623L..37M}.
While the spectroscopic profiles are once again mainly dictated by the light-curve decline rate parameter \citep{2009PhDT.......228H, 2014MNRAS.437..338C}, the variations are not completely described by one parameter.
Both photometric and spectroscopic studies have shown that longer wavelength regions indeed provide independent information from the traditional $B$ and $V$ bands \citep[e.g.,][]{2011ApJ...731..120M, 2019PASP..131a4002H}.
Cosmological studies have also taken advantage of the rest-frame $i$-band to provide independent confirmation of dark energy \citep{2005A&A...437..789N, 2009ApJ...704.1036F}.

The $i$-band light-curves of SNe~Ia generally have more complex shapes than those in the $B$ and $V$ bands.
Most have a secondary maximum occurring approximately $20-30$ days after the primary maximum.
\citet{2006ApJ...649..939K} modeled the wavelength region and determined the phenomenon to be the manifestation of the ionization evolution of the synthesized iron-peak elements.
The overall shape is still largely dictated by the decline rate parameter $\Delta m_{15} (B)$ \citep{2001AJ....122.1616K}.
A fast-declining SN~Ia tends to have a weaker secondary maximum occurring earlier and thus closer to the primary maximum.
In the extreme case of a fast-declining subluminous 91bg-like SN~Ia, the secondary maximum merges completely with the primary one and is absent.

Large scatter is also seen in the strength of the secondary maximum and the depth of the trough between the two maxima \citep{2010AJ....139..120F}.
The $i$-band region is also effective for identifying peculiar subtypes of SNe~Ia \citep{2014ApJ...795..142G}.
\citet{2020ApJ...895L...3A} concluded that all normal SNe~Ia have the $i$-band primary maxima taking place before the $B$-band maxima.
On the other hand, the peculiar 02cx-like, 03fg-like, and 91bg-like SNe~Ia largely have their primary $i$-band maxima occurring after those in $B$-band.

In this work, we focus on an interesting feature in the $i$-band light-curve that emerges a few days past the $i$-band primary maximum.
There appears to be an abrupt change in the curvature or concavity around this epoch for most $i$-band light-curves. 
In some cases, this flattens the light-curve within a window of several days.
In the extreme, the abrupt change in curvature can appear as a strong downward concave shape or a ``kink.''
This sometimes subtle feature is only apparent in a data set that is of high signal-to-noise ratio (S/N) ratio and rapid cadence, such as that obtained by the Carnegie Supernova Project \citep[CSP;][]{2006PASP..118....2H, 2019PASP..131a4001P}.
In some figures previously published by the CSP, these features can easily be detected by eye without any statistical treatment: e.g., Figure~1 of \citet{2010AJ....139..120F} and Figure~8 of \citet{2011AJ....141...19B}.
The feature is generally only present in the $i$-band.
Also note that this feature is not the secondary maximum and occurs before the minimum that is sandwiched between the two maxima.

Here, we investigate this change in concavity in the $i$-band in detail. 
The source of light-curves, sample selection, and characteristics are detailed in Section~\ref{sec:sample}. 
The method for quantifying the feature and tests for the robustness of the measurements are presented in Section~\ref{sec:methods}. 
The resulting measurements are plotted against several known SN~Ia parameters in Section~\ref{sec:results}. 
The conclusions are then summarized in Section~\ref{sec:conclusion}.

\section{The Light-curve Sample}
\label{sec:sample}

The light-curve sample is entirely drawn from the CSP.
Two phases of the project: CSP-I \citep{2006PASP..118....2H} and CSP-II \citep{2019PASP..131a4001P,2019PASP..131a4002H}, were both NSF-supported and conducted from $2004-2009$ and $2011-2015$, respectively.
All light-curves were obtained using a single telescope, the 1-m Swope Telescope at Las Campanas Observatory (LCO), with a well-understood photometric system.
The uniformity of the data allows for the study of subtle light-curve features such as the one in this work.

\subsection{Data Source}

During most of CSP-I, the Swope Telescope was used to obtain both optical $BVugri$ and NIR $YJH$ light-curves with the SITe3 CCD and RetroCam imagers, respectively. 
During CSP-II, the Swope Telescope was dedicated to obtaining optical light-curves with the SITe3 imager, as well as the e2v imager following an upgrade.
Noting interesting $i$-band light-curve features from observations during CSP-I, special care was taken to obtain nightly data points around the $i$-band primary and secondary maxima during CSP-II.
Furthermore, CSP-II aimed to follow SNe~Ia in the Hubble flow discovered by untargeted surveys to build an unbiased sample with minimal uncertainties from host galaxy peculiar velocities.

The photometry is taken from previous data releases of CSP-I \citep{2010AJ....139..519C, 2011AJ....142..156S, 2017AJ....154..211K} as well as soon-to-be-published light-curves of CSP-II.
The light-curves have been K-corrected using the SuperNovae in the object oriented {\sc Python} package \citep[SNooPy;][]{2011AJ....141...19B} utilizing the updated spectral template of \citet{2007ApJ...663.1187H}.
Since this work is focused on the light-curve shape of a single monochromatic band, the extinction corrections do not have significant effects on the analysis.

\subsection{Light-curve Interpolation}
\label{subsec:GP}

To compare the $i$-band light-curves in a consistent manner, we elected to interpolate the light-curves via Gaussian Process (GP) using the {\sc Python} package {\sc GPy}\footnote{{GPy}: A Gaussian process framework in {\sc Python}. See \url{http://github.com/SheffieldML/GPy}.} with the set up described in \citet{2019MNRAS.488.4239P}. 
The first approximations of the length scale and variance hyperparameters were set to be the cadence and the standard deviation of said separations, respectively.
The cadence is defined here as the average rest-frame separation of successive photometric points in days.
Approximately 10\% of the sample required minor adjustments to the above hyperparameters in order to obtain a good interpolation.

The mean of the GP functions was then taken as the interpolated $i$-band light-curve and was used to measure the timing and the magnitude of the primary maximum.
The interpolation is limited to the phase range between the first observed point and the minimum or shoulder between the primary and secondary maxima. 
If no secondary maximum exists in a light-curve, the interpolation stops at around 20 days after the primary maximum.
The light-curve phases quoted in the remaining text are stated relative to the $i$-band primary maximum and are corrected for time dilation.

\subsection{Sample Selection}

Our sample selection is mainly based on the observed phase range and cadence.
But first, we removed members of peculiar SN~Ia subgroups: 02cx-like \citep{2003PASP..115..453L}, 02ic-like \citep{2003Natur.424..651H}, 03fg-like \citep{2006Natur.443..308H}, and 06bt-like \citep{2010ApJ...708.1748F}.
Those SNe~Ia identified as 91T-like and 91bg-like via their classification spectra were kept in the sample, since these classifications were uncertain and there are indications that they may be part of the normal population \citep[e.g.,][]{1995ApJ...455L.147N}.
Given that our aim is to characterize a light-curve feature that appears soon after the $i$-band primary maximum, we also limit the sample to those objects for which the primary maximum is observed. 
In other words, there must exist at least one photometric point prior to the time of primary maximum.

Following a visual examination of our sample, the $i$-band light-curve feature, if present, occurs no earlier than 0.5~d and no later than 7.5~d past the primary maximum. 
In order to adequately characterize the feature, the light-curves need to be well sampled in this region, since the abrupt change in curvature lasts only for a few days.
Thus, we first compile a ``Gold'' sample consisting of objects with nightly observations or a rest-frame cadence of $<1$~d between $0.5-7.5$~d past maximum.
Furthermore, the maximum allowed gap between successive photometric points in the light-curve is one day.
The Gold sample includes 19 SNe~Ia. These are our best sampled light-curves.

Second, we construct a ``Silver'' sample with a slightly relaxed requirement for rest-frame cadence: $1-1.5$~d in the phase range of $0.5-7.5$~d past maximum.
The choice of the 1.5~d cadence is based on whether the light-curve feature is detectable through worsening cadence and will be further discussed in Section~\ref{sec:methods}.
Additionally, the maximum gap between successive photometric points cannot be larger than 3~d.
The Silver sample adds 35 SNe~Ia to our analyses, bringing the total to 54 SNe~Ia. 
These objects are listed in Table~\ref{tab:sample}.

\subsection{Sample Characteristics}

\begin{figure*}
\centering
\includegraphics[width=\textwidth]{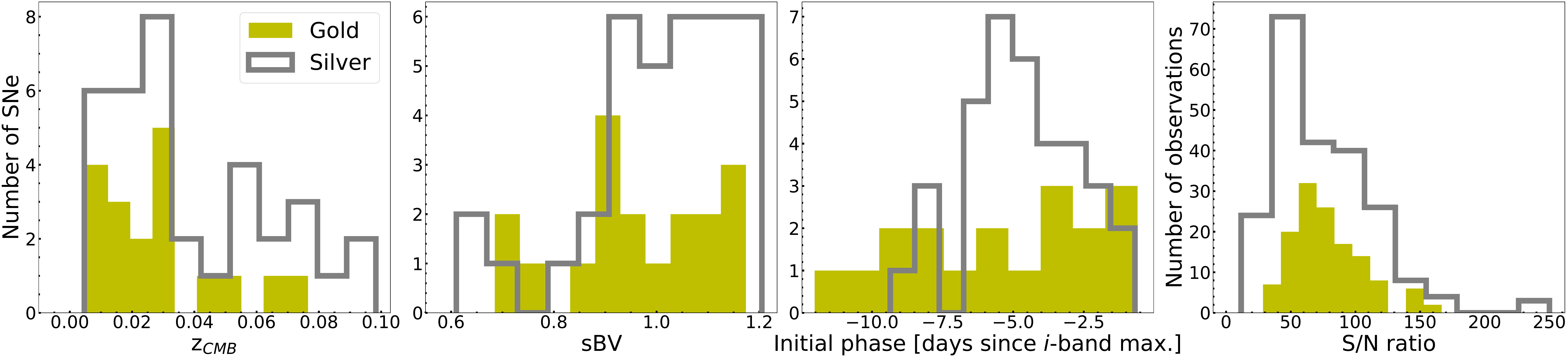}
\caption{The sample distribution in CMB redshift, color-stretch parameter \sBV, the rest-frame phase relative to $i$-band maximum of the first observed point, and the S/N ratio of the observations. 
The Gold and Silver samples are represented in filled gold and open silver bars, respectively.}
\label{fig:distri}
\end{figure*}

Some basic properties of our light-curve samples are presented in Figure~\ref{fig:distri}.
The distribution of redshifts shows that most SNe~Ia are in the Hubble flow ($z\gtrsim0.02$) where distance uncertainties due to the host galaxy peculiar velocities are small compared to the Universal expansion.
The redshifts shown have been corrected to the rest frame of the cosmic microwave background.

The primary parameter of normal SNe~Ia is their light-curve decline rate which correlates with their peak absolute magnitude \citep[e.g.,][]{1993ApJ...413L.105P}.
Here, we use the primary parameter \sBV\ described above.
For our sample of SNe~Ia, \sBV\ parameters were determined via SNooPy multi-band light-curve fits.
The Gold and Silver samples have similar distributions in \sBV, covering a wide range and thus, both intrinsically faint and bright SNe~Ia. 

Approximately 90\% of the SNe~Ia in the Gold and Silver samples were drawn from CSP-II, mainly due to the cadence selection criterion.
Consequently, around three quarters of the SNe~Ia were drawn from untargeted searches where the host galaxies were not pre-selected.
Thus, our samples are not significantly biased toward more massive host galaxies, even though they only include nearby objects ($z < 0.1$). 

The two samples have good time coverage and cadence by design.
As the $i$-band maximum occurs before that in the $B$-band, follow-up observations must start early.
Figure~\ref{fig:distri} shows the rest-frame phase of the first photometric point of each SN~Ia, and a large fraction of them are well covered at early phases.
The vast majority of the observations are also of high precision, with 76\% of the photometric points with S/N ratios of 50 or higher.

\section{Quantifying the light-curve feature}
\label{sec:methods}

The $i$-band light-curve feature we are attempting to quantify appears as an abrupt change in the curvature or concavity between the primary and the secondary maxima.
When present, it manifests as a subtle flattening in the light-curve or a stronger downward concave shape, a kind of ``kink.''
Since we are dealing with a change in curvature, the time derivatives of the light-curve are naturally the most effective tool for detecting such a feature.
This approach has been adopted before to study light-curves of core-collapse SNe \citep{2019MNRAS.488.4239P}.

\subsection{The measurements}
\label{subsec:measurements}

\begin{figure*}
\centering
\includegraphics[width=\textwidth]{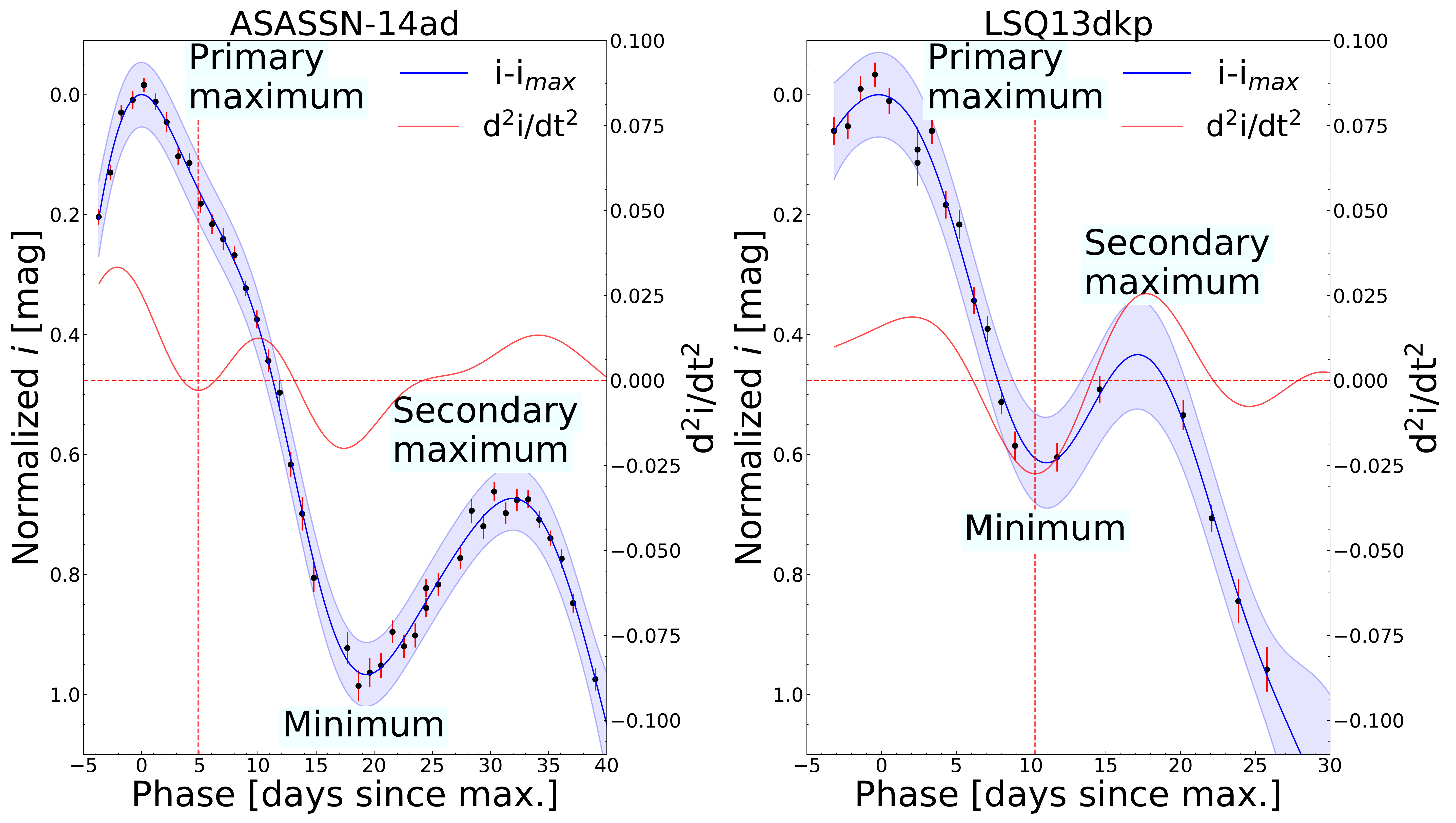}
\caption{Examples of SN~Ia $i$-band light-curves and their derivatives. The primary and secondary maxima, as well as the minimum in between, are noted. The GP interpolations are shown as blue curves (mean) and bands (confidence region) through the data points of two SNe~Ia. 
The second derivative of the interpolation with respect to time is shown in red. The red horizontal dashed line indicates the zero in the derivative space and the red vertical dashed line indicates the first minimum of the second derivative after peak. 
The left and right panels show examples of a detection and a non-detection of the light-curve feature, respectively.}
\label{fig:example}
\end{figure*}

We use the GP interpolated light-curves described in Section~\ref{subsec:GP} to determine the time derivatives. 
The second derivative was used to characterize the curvature of the light-curves.
Specifically, we locate the first local minimum of the second derivative after the primary peak where the change of curvature is detected, if such a feature exists.
Figure~\ref{fig:example} shows two examples of $i$-band light-curves and their second derivatives.
In the left panel, ASASSN-14ad is shown as an example of a modest light-curve flattening feature.
The local minimum of the second derivative successfully locates the center of the feature, at approximately 5~days past maximum. 
Note that the light-curve minimum between the primary and secondary maxima also produces a local minimum in the second derivative curve.
If no second derivative local minimum is detected between the primary maximum and minimum (i.e., the first local minimum in the second derivative is at the light-curve minimum), the case is deemed a non-detection for the light-curve feature.
The right panel of Figure~\ref{fig:example} shows LSQ13dkp as an example where no local minimum in the second derivative exists before the light-curve minimum.

\begin{figure}
\centering
\includegraphics[width=\columnwidth]{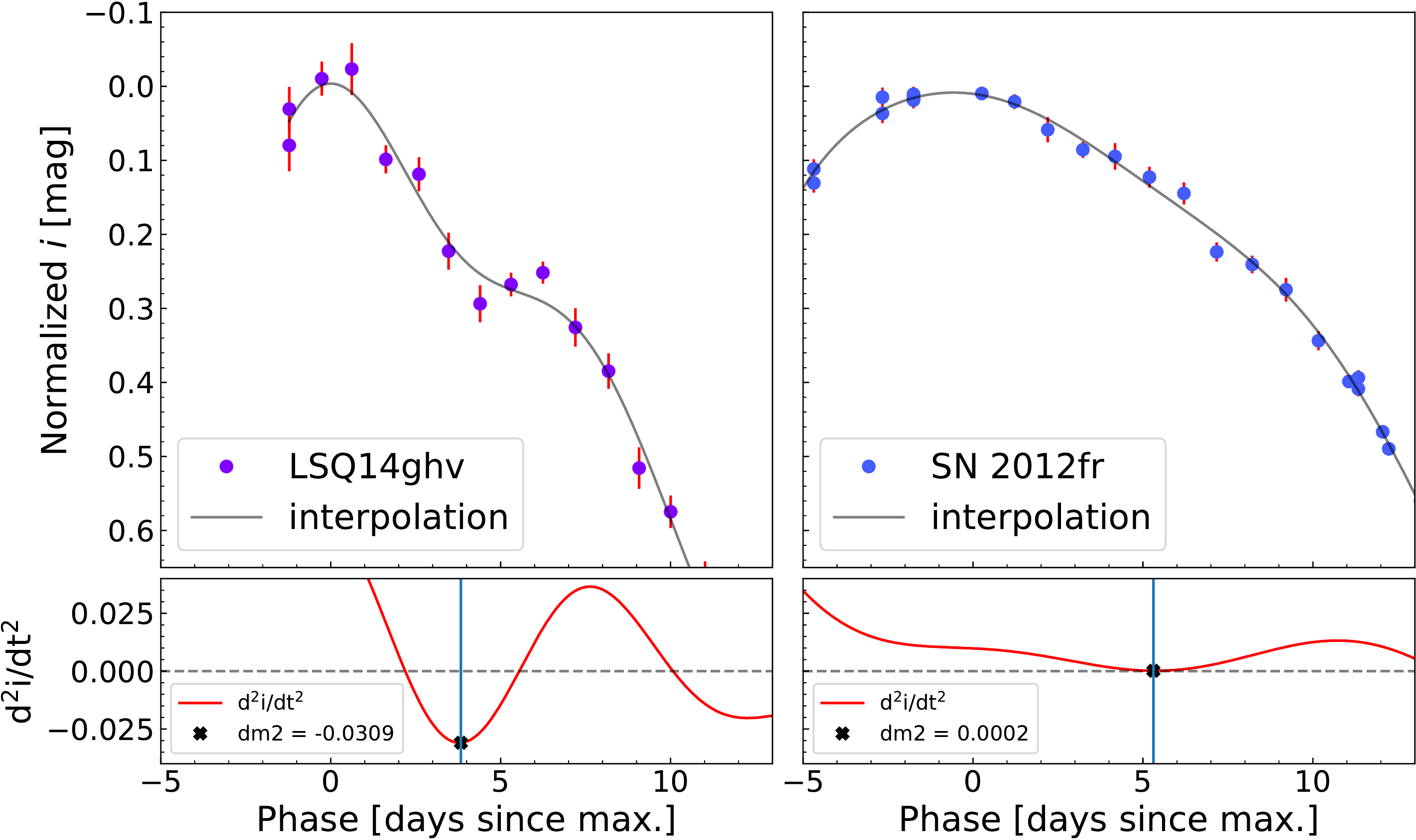}
\caption{Examples of negative and positive \dm. Left panel: LSQ14ghv showing a large negative \dm\ and a strong downward concave feature or ``kink.''
Right panel: SN~2012fr showing a small positive \dm\ and a subtle flattened feature. 
The bottom panels show the behavior of the respective second derivatives.
Blue vertical lines mark the phases of \tdm.
}
\label{fig:parameterdef}
\end{figure}

This method provides two parameters for a detected $i$-band light-curve feature:
\begin{enumerate}
    \item The timing of the feature, \tdm, defined as the rest-frame phase relative to the primary maximum when the second derivative reaches a local minimum.
    \item The strength of the curvature and the direction of concavity, \dm, defined as the value and the sign of the second derivative measured at \tdm, respectively.
\end{enumerate}
A strong downward concave feature would produce a large negative \dm\ (left panel of Figure~\ref{fig:parameterdef}).
A more subtle flattened feature would have a small negative and or a small positive \dm\ (right panel of Figure~\ref{fig:parameterdef}).

\begin{figure}
\centering
\includegraphics[width=\columnwidth]{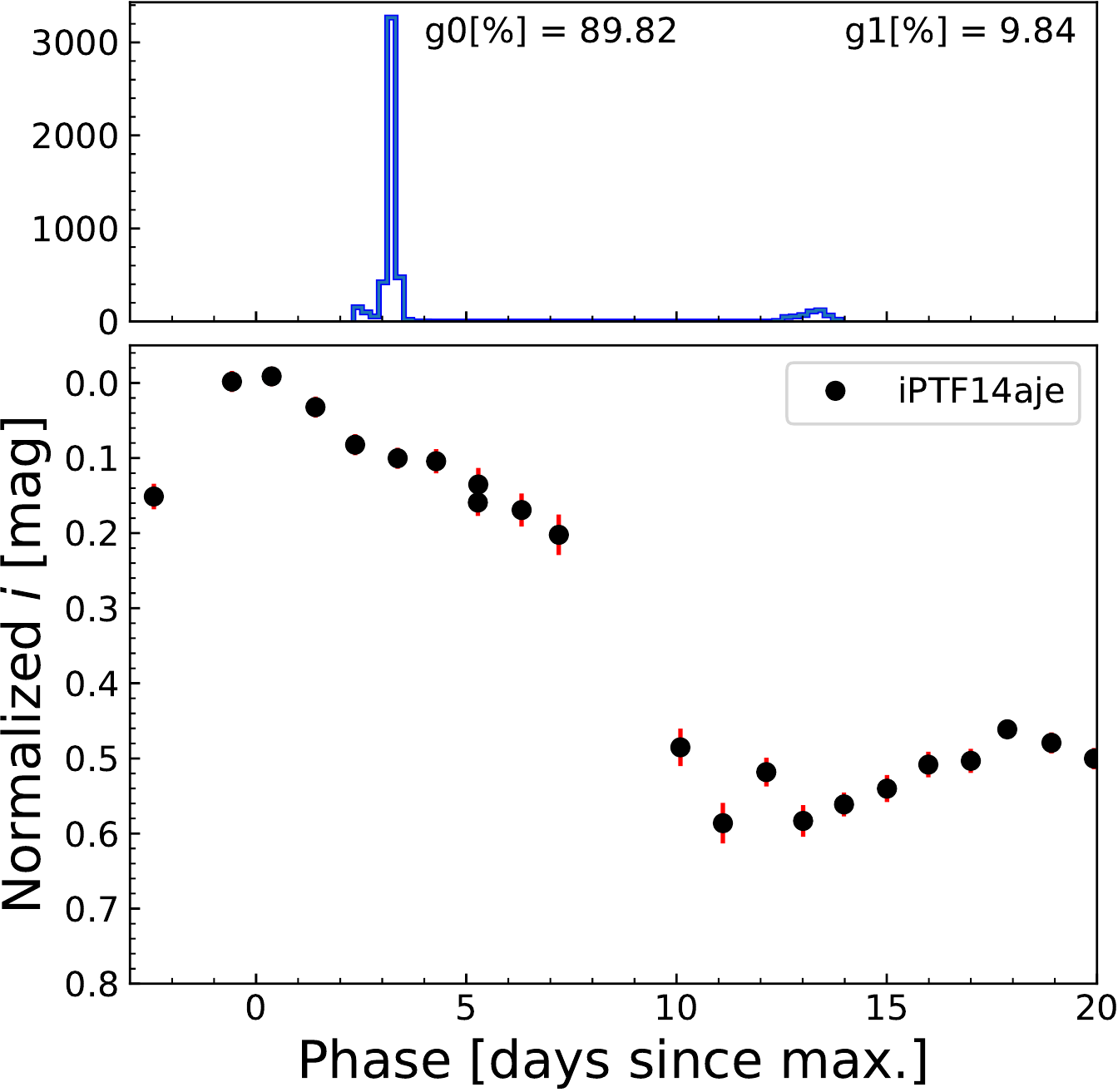}
\caption{An example of the phase measurements of the first minimum of the second derivative after peak through a Monte Carlo process.
This exercise examines the effects of uncertainties in the GP interpolation process and estimates the corresponding uncertainties for \tdm\ and \dm\ measurements.
The top panel shows the histogram of the resulting phase values, and the bottom panel shows the light-curve itself with a clear light-curve feature.}
\label{fig:hist}
\end{figure}

To assess if the detection is robust against uncertainties in the interpolation process, we obtained 5000 Monte Carlo realizations of the resultant GP model.
The second derivative curve was then computed for each iteration. 
The resulting distribution of phases at which the second derivative shows the first minimum after peak is used to calculate the rate of detection of the light-curve feature. 
The distribution is almost always bimodal as the light-curve feature is detected in some realizations (detections) and the light-curve minimum between the primary and secondary maxima is detected for others (non-detections, see Figure~\ref{fig:hist}).
The group of phases with the higher percentage of occurrence is used to determine if there is a detection or non-detection of the light-curve feature. 
If the percentages of two phase groups are similar, the result is considered to be the one closer to that obtained using the mean GP interpolation. 
The uncertainties for \tdm\ and \dm\ were then taken from the standard deviation of the respective distributions.
The \tdm\ and \dm\ values and their uncertainties are presented in Table~\ref{tab:sample}.

\begin{figure*}
\centering
\includegraphics[width=\textwidth]{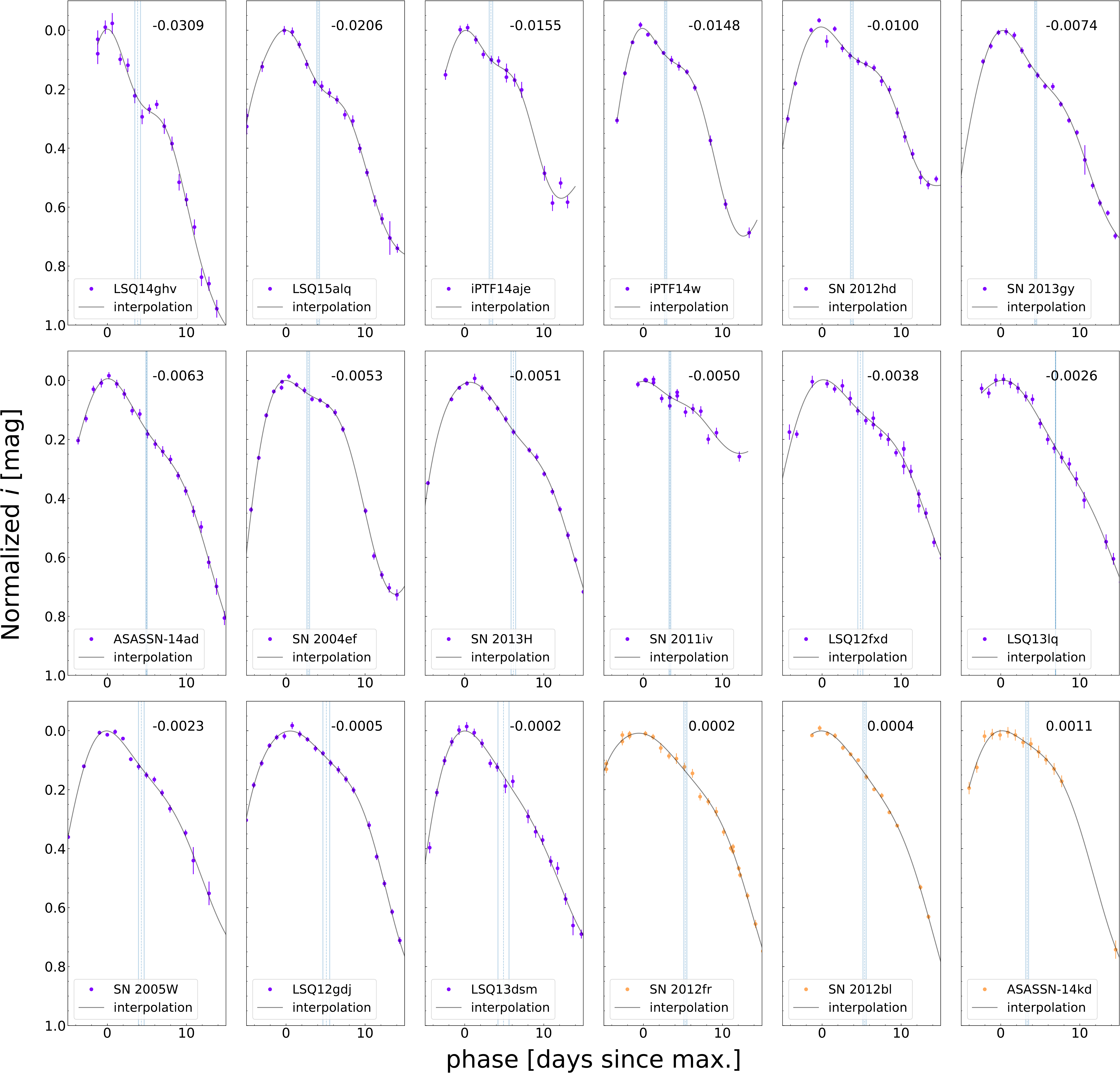}
\caption{The Gold sample $i$-band observed and interpolated light-curves.
The light-curves are normalized to maximum light.
The panels are sorted by \dm, noted in the upper right corner of each panel. 
SNe~Ia with negative and positive \dm\ are plotted in purple and orange, respectively. 
The blue vertical dashed line in each panel locates the \tdm, while the blue vertical solid lines denote the 1-$\sigma$ uncertainty range.
}
\label{fig:platinumsorted}
\end{figure*}

\begin{figure*}
\centering
\includegraphics[width=\textwidth]{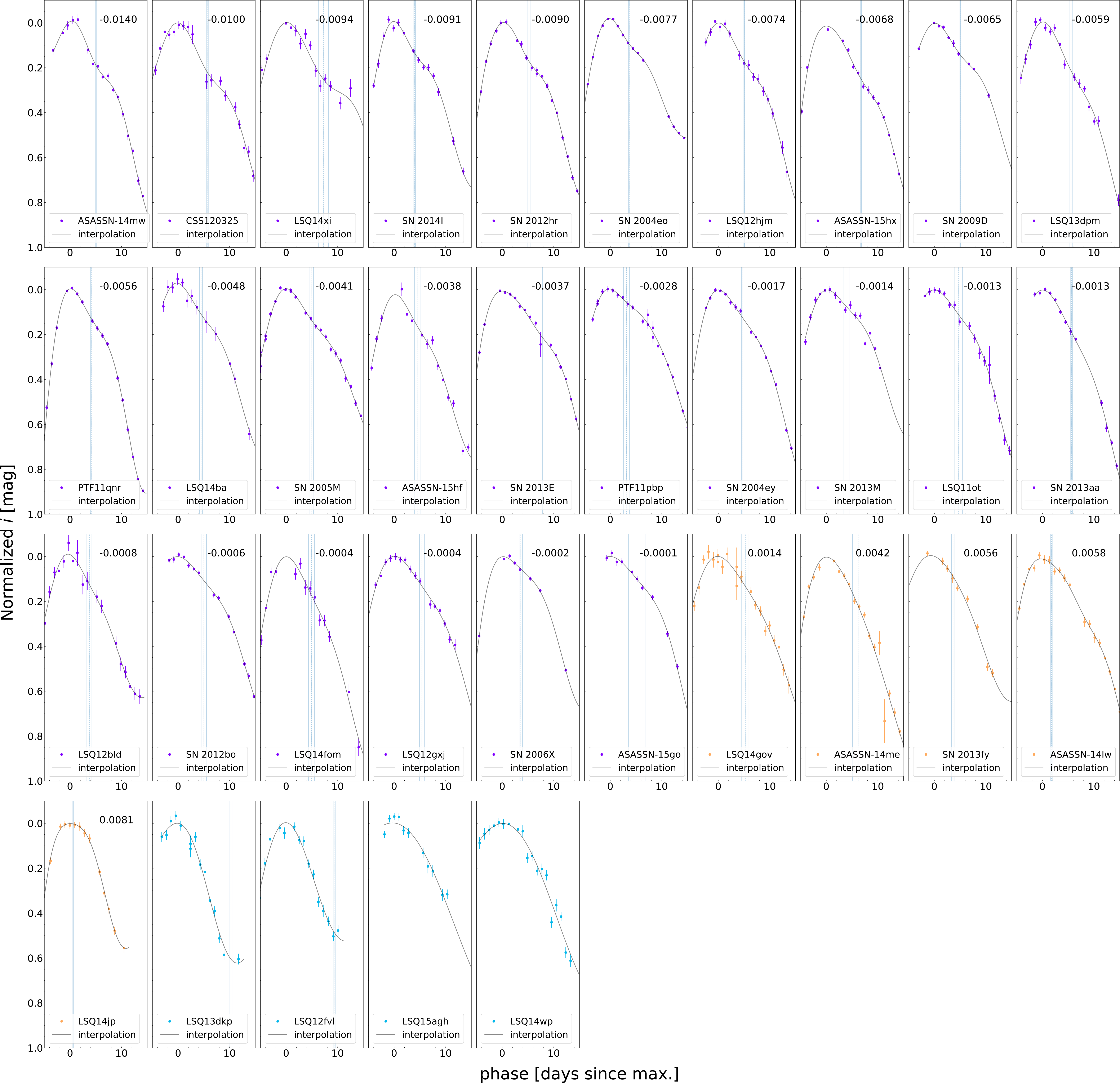}
\caption{The Silver sample $i$-band observed and interpolated light-curves. 
The figure is presented in the same way as Figure~\ref{fig:platinumsorted} with the addition of SN~Ia with a non-detections shown in cyan.
}
\label{fig:goldsorted}
\end{figure*}

The $i$-band light-curves of the SNe~Ia in the Gold and Silver samples as well as their mean GP interpolations are presented in Figures~\ref{fig:platinumsorted} and \ref{fig:goldsorted}, respectively. 
SNe~Ia are sorted by their value of \dm.
A more negative value of \dm\ indicates a stronger feature.
There appears to be a full range of strengths without clear groups of distinct light-curve shapes.
Only five SNe~Ia, one in the Gold (although see Section \ref{subsec:snratio}) and four in the Silver sample, have non-detections.
A light-curve feature with either weak or strong downward curvature is clearly prevalent in most SNe~Ia.

\subsection{The effects of S/N ratio and cadence}
\label{subsec:snratio}
Whether a light-curve feature can be detected is dependent on the S/N ratio of the observations.
We conducted multiple tests in an attempt to locate the S/N threshold to detect the features.
Several light-curves from the Gold sample with the light-curve features at various strengths were adopted for the tests.
The mean GP interpolations were taken as the idealized light-curve with an infinite S/N ratio.
Random noise was then added to the idealized light-curve simulating observations with a range of precision.

The threshold depends on the strength of the feature and the placement of high S/N ratio points.
A strong feature with a large negative \dm\ generally does not require high S/N to be detected.
A modest strength feature such as that of SN~2013gy (\dm$=-0.0074$~\magdsq, see Figure~\ref{fig:platinumsorted}) requires a light-curve with a median S/N ratio of higher than 40 or a magnitude error lower than $\sim0.03$~mag.
All of our light-curves satisfy this threshold except for that of LSQ12aor which yielded a non-detection.
LSQ12aor is thus excluded from the remaining analyses.
Other non-detections also have slightly lower S/N ratio than the average of the whole sample.
Since all these light-curves have median S/N ratios higher than the derived threshold of 40, they are not likely to conceal strong light-curve features and will be grouped with other SNe~Ia showing weak features in the remaining analyses.
The distribution of the precision of our individual light-curve points is shown in the right panel of Figure~\ref{fig:distri}.

How often the light-curve points are sampled can also dictate the detection and parameter measurements.
To prevent the effects of the cadence choice from entering into our results, we performed cadence tests in which random points were removed from light-curves in the Gold sample to generate simulated light-curves with degraded cadence ranging from 1 to 3.5 days. 
The parameters \tdm\ and \dm\ were then measured for each simulated light-curve and their distributions were examined. 

For light-curves with a cadence of $1-1.5$~days, more than half have resulting \tdm\ and \dm\ measurements within 1-$\sigma$ uncertainty of the mean of the distribution. 
This value increases by up to 10\% when considering values within 2-$\sigma$ (i.e., if 60\% of the values lie within 1-$\sigma$, then 70\% of the values lie within 2-$\sigma$).
In this cadence range, the rate of detection is also greater than 75\% when a light-curve feature is present compared to a rate of detection of as low as 50\% for a cadence of 2.5~days.
We therefore selected the 1.5~day cadence limit for the Silver sample in an attempt to increase the sample size without significantly introducing the effects of cadence.
We note that for a handful of cases, especially those with large negative \dm\ values, the removal of a few critical points could alter the resulting \dm, even with a rapid cadence.
This effect is accounted for in our method and is the reason for the larger uncertainties of \dm\ for those light-curves with large negative \dm\ values.

\begin{table*}
\caption{The SN~Ia sample, their sources, redshifts, \dm\, \tdm\ and $\Delta \mu$ measurements.} 
\label{tab:sample}
\begin{threeparttable}
\begin{tabular}{lccccccr}
\hline
SN Name & Sample & CSP\tnote{a} & Untargeted\tnote{b} & z$_{CMB}$ & \dm\ [mag~d$^{-2}$] & \tdm\ [d] & $\Delta \mu$ [mag]\tnote{c}\\
\hline
ASASSN-14ad             &  Gold   &  II & Yes & 0.027   &   $-$0.0063(0.0013) &  5.0(0.1)  &    0.185(0.185)   \\
ASASSN-14kd             &  Gold   &  II & Yes & 0.023   &      0.0011(0.0016) &  3.3(0.2)  & $-$0.266(0.202)   \\
ASASSN-14lw             &  Silver &  II & Yes & 0.020   &      0.0058(0.0012) &  1.8(0.2)  &    0.312(0.195)   \\
ASASSN-14me             &  Silver &  II & Yes & 0.017   &      0.0042(0.0008) &  6.2(1.1)  & $-$0.010(0.216)   \\   
ASASSN-14mw             &  Silver &  II & Yes & 0.027   &   $-$0.0140(0.0051) &  5.0(0.1)  & $-$0.002(0.177)   \\       
ASASSN-15go             &  Silver &  II & Yes & 0.019   &   $-$0.0001(0.0009) &  5.2(1.6)  &    0.076(0.215)   \\ 
ASASSN-15hf             &  Silver &  II & Yes & 0.007   &   $-$0.0038(0.0077) &  4.5(0.6)  &    0.311(0.389)   \\
ASASSN-15hx             &  Silver &  II & Yes & 0.009   &   $-$0.0068(0.0012) &  6.7(0.1)  & $-$0.463(0.318)   \\ 
CSS120325:123816-150632 &  Silver &  II & Yes & 0.098   &   $-$0.0100(0.0083) &  5.7(0.2)  & $-$0.061(0.152)   \\  
LSQ11ot                 &  Silver &  II & Yes & 0.027   &   $-$0.0013(0.0020) &  4.7(0.7)  &    0.162(0.194)   \\       
LSQ12aor                &  Gold   &  II & Yes & 0.094   &   $\cdots$          &  $\cdots$  &    0.059(0.182)   \\  
LSQ12bld                &  Silver &  II & Yes & 0.084   &   $-$0.0008(0.0034) &  3.8(0.5)  & $-$0.006(0.163)   \\   
LSQ12fvl                &  Silver &  II & Yes & 0.056   &   $\cdots$          &  $\cdots$  &    0.127(0.166)   \\      
LSQ12fxd                &  Gold   &  II & Yes & 0.031   &   $-$0.0038(0.0011) &  4.8(0.3)  & $-$0.102(0.179)   \\      
LSQ12gdj                &  Gold   &  II & Yes & 0.029   &   $-$0.0005(0.0011) &  5.1(0.4)  & $-$0.114(0.182)   \\  
LSQ12gxj                &  Silver &  II & Yes & 0.034   &   $-$0.0004(0.0011) &  5.4(0.5)  &    0.416(0.241)   \\  
LSQ12hjm                &  Silver &  II & Yes & 0.071   &   $-$0.0074(0.0009) &  5.0(0.1)  &    0.112(0.164)   \\    
LSQ13lq                 &  Gold   &  II & Yes & 0.076   &   $-$0.0026(0.0009) &  6.9(0.0)  & $-$0.213(0.163)   \\
LSQ13dkp                &  Silver &  II & Yes & 0.068   &   $\cdots$          &  $\cdots$  &    0.206(0.164)   \\   
LSQ13dpm                &  Silver &  II & Yes & 0.052   &   $-$0.0059(0.0022) &  5.6(0.2)  & $-$0.077(0.167)   \\   
LSQ13dsm                &  Gold   &  II & Yes & 0.042   &   $-$0.0002(0.0014) &  4.9(0.7)  & $-$0.056(0.171)   \\  
LSQ14ba                 &  Silver &  II & Yes & 0.079   &   $-$0.0048(0.0034) &  4.5(0.3)  &    0.051(0.163)   \\  
LSQ14jp                 &  Silver &  II & Yes & 0.046   &      0.0081(0.0017) &  0.6(0.1)  & $-$0.012(0.178)   \\
LSQ14wp                 &  Silver &  II & Yes & 0.071   &   $\cdots$          &  $\cdots$  & $-$0.079(0.164)   \\   
LSQ14xi                 &  Silver &  II & Yes & 0.052   &   $-$0.0094(0.0034) &  7.2(1.0)  & $-$0.040(0.167)   \\ 
LSQ14fom                &  Silver &  II & Yes & 0.055   &   $-$0.0004(0.0037) &  4.9(0.6)  & $-$0.128(0.166)   \\  
LSQ14ghv                &  Gold   &  II & Yes & 0.066   &   $-$0.0309(0.0131) &  3.8(0.4)  &    0.123(0.164)   \\   
LSQ14gov                &  Silver &  II & Yes & 0.089   &      0.0014(0.0010) &  5.3(0.7)  &    0.080(0.153)   \\
LSQ15agh                &  Silver &  II & Yes & 0.061   &   $\cdots$          &  $\cdots$  &    0.024(0.165)   \\  
LSQ15alq                &  Gold   &  II & Yes & 0.048   &   $-$0.0206(0.0060) &  4.1(0.2)  & $-$0.168(0.168)   \\   
PTF11pbp                &  Silver &  II & Yes & 0.028   &   $-$0.0028(0.0017) &  3.1(0.6)  &    0.001(0.184)   \\       
PTF11qnr                &  Silver &  II & Yes & 0.015   &   $-$0.0056(0.0013) &  4.1(0.1)  &    0.033(0.233)   \\    
iPTF14w                 &  Gold   &  II & Yes & 0.020   &   $-$0.0148(0.0012) &  2.8(0.1)  &    0.175(0.204)   \\
iPTF14aje               &  Gold   &  II & Yes & 0.028   &   $-$0.0155(0.0081) &  3.3(0.2)  & $-$0.119(0.192)   \\   
SN~2004ef               &  Gold   &  I  & No  & 0.030   &   $-$0.0053(0.0008) &  2.8(0.2)  &    0.119(0.181)   \\   
SN~2004eo               &  Silver &  I  & No  & 0.015   &   $-$0.0077(0.0005) &  3.7(0.1)  &    0.082(0.235)   \\    
SN~2004ey               &  Silver &  I  & No  & 0.015   &   $-$0.0017(0.0006) &  4.6(0.1)  &    0.188(0.236)   \\     
SN~2005M                &  Silver &  I  & No  & 0.025   &   $-$0.0041(0.0008) &  5.0(0.4)  &    0.026(0.188)   \\      
SN~2005W                &  Gold   &  I  & No  & 0.008   &   $-$0.0023(0.0014) &  4.3(0.4)  &    0.154(0.371)   \\ 
SN~2006X                &  Silver &  I  & No  & 0.006   &   $-$0.0002(0.0022) &  3.6(0.3)  & $-$1.191(0.432)   \\  
SN~2009D                &  Silver &  I  & No  & 0.025   &   $-$0.0065(0.0013) &  5.0(0.1)  & $-$0.001(0.190)   \\     
SN~2011iv               &  Gold   &  II & -   & 0.006   &   $-$0.0050(0.0051) &  3.4(0.1)  & $-$0.236(0.440)   \\ 
SN~2012bl               &  Gold   &  II & No  & 0.018   &      0.0004(0.0003) &  5.4(0.2)  &    0.149(0.205)   \\ 
SN~2012bo               &  Silver &  II & No  & 0.026   &   $-$0.0006(0.0012) &  5.0(0.5)  &    0.080(0.186)   \\  
SN~2012fr               &  Gold   &  II & No  & 0.005   &      0.0002(0.0005) &  5.3(0.2)  &    0.127(0.519)   \\
SN~2012hd               &  Gold   &  II & No  & 0.011   &   $-$0.0100(0.0047) &  3.8(0.2)  &    0.157(0.278)   \\ 
SN~2012hr               &  Silver &  II & No  & 0.008   &   $-$0.0090(0.0022) &  5.2(0.2)  &    0.198(0.360)   \\  
SN~2013E                &  Silver &  II & No  & 0.010   &   $-$0.0037(0.0007) &  7.1(0.8)  & $-$0.291(0.290)   \\
SN~2013H                &  Gold   &  II & No  & 0.016   &   $-$0.0051(0.0013) &  6.2(0.3)  &    0.001(0.226)   \\
SN~2013M                &  Silver &  II & No  & 0.036   &   $-$0.0014(0.0019) &  4.0(0.6)  & $-$0.254(0.175)   \\
SN~2013aa               &  Silver &  II & No  & 0.005   &   $-$0.0013(0.0014) &  5.7(0.2)  & $-$0.930(0.566)   \\      
SN~2013fy               &  Silver &  II & No  & 0.030   &      0.0056(0.0013) &  3.6(0.3)  & $-$0.021(0.181)   \\  
SN~2013gy               &  Gold   &  II & Yes & 0.013   &   $-$0.0074(0.0016) &  4.4(0.1)  & $-$0.074(0.246)   \\      
SN~2014I                &  Silver &  II & Yes & 0.030   &   $-$0.0091(0.0023) &  4.0(0.1)  & $-$0.100(0.172)   \\ 
\end{tabular}
\end{threeparttable}
\begin{tablenotes}
\small
\item a -- This column identifies the CSP follow-up project: CSP-I \citep{2006PASP..118....2H} or CSP-II \citep{2019PASP..131a4001P}.
\item b -- This column identifies whether the SNe~Ia was discovered by an untargeted transient search survey.
\item c -- These results will be presented in Burns et al. \emph{in prep}. 
\end{tablenotes}
\end{table*}

\section{Results}
\label{sec:results}

In previous sections we have introduced and measured two light-curve parameters based solely on the monochromatic $i$-band light-curves of SNe~Ia.
The method is completely independent of those used to measure the standard light-curve parameters for SNe~Ia.
The natural next step is to explore whether these new $i$-band parameters provide independent information about the SNe~Ia.
If they do, can they provide insights into their physical origins and/or improve standardization for SN~Ia cosmology as a secondary parameter?
Such possibilities are explored in this section.

\subsection{Light-curve parameters \tdm\ and \dm}
\label{subsec:tdmdm}

Our two main parameters \tdm\ and \dm\ measure the timing (relative to $i$-band maximum) and the strength of the concavity of the $i$-band light-curve feature, respectively.

\begin{figure}
\includegraphics[width=\columnwidth]{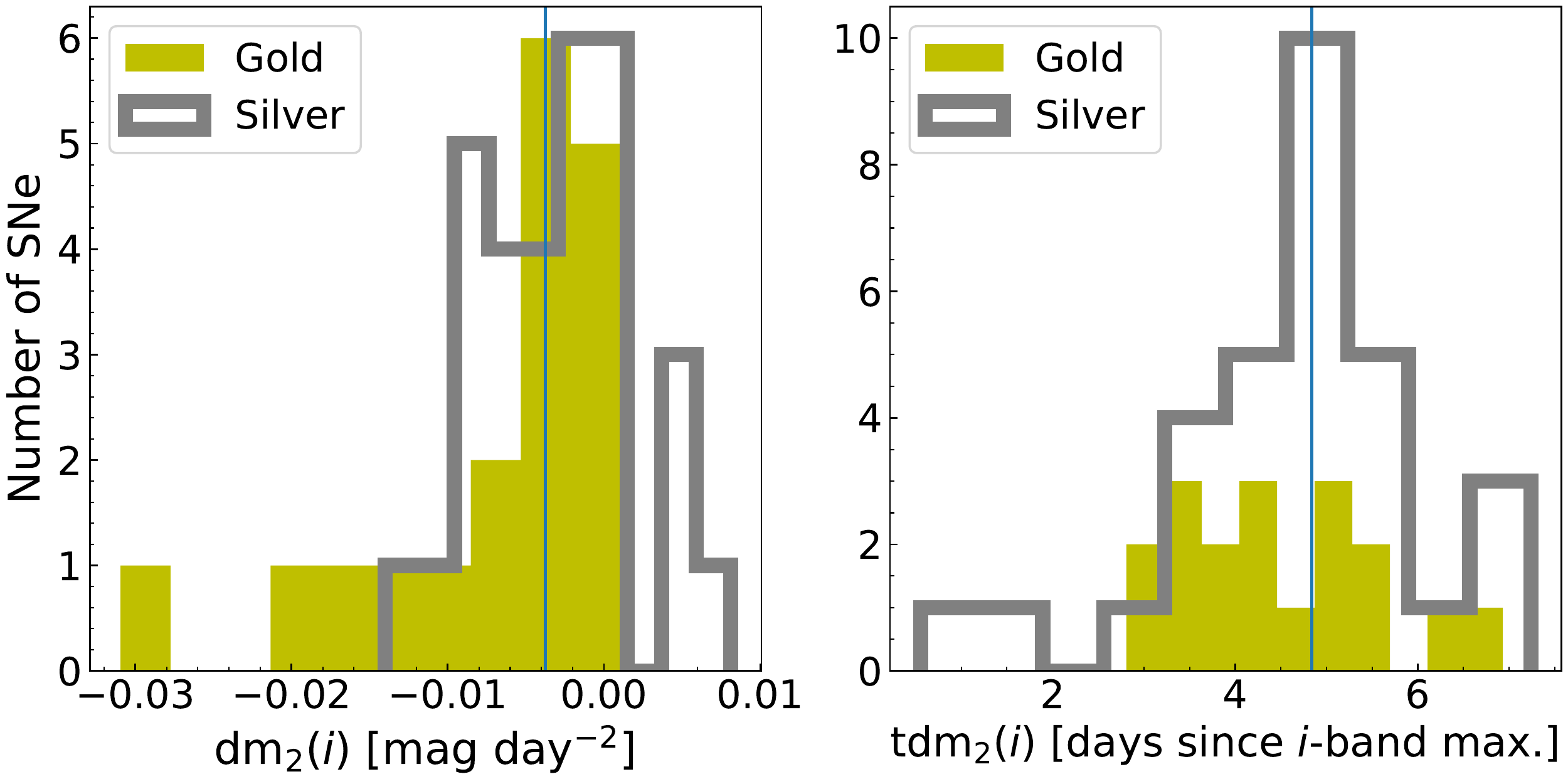}
\caption{
The distributions of the \dm\ (left panel) and \tdm\ (right panel) measurements.
The vertical blue lines show the median values for the joint Gold plus Silver sample. The Gold sample is presented in yellow, while the Silver sample is presented in gray.
}
\label{fig:dmtdm_distrib}
\end{figure}

The distributions of these two parameters in the Gold and Silver samples are presented in Figure~\ref{fig:dmtdm_distrib}.
The values of \tdm\ span from just past $i$-band maximum to no later than 7.5~d past with a median value of 4.8~d. 
All the second derivative minima measured past 7.5~d originate from the light-curve minimum between the primary and secondary maxima and therefore constitute non-detections.
The values of \dm\ span from large negative values ($\sim-0.03$) to small positive values ($0.01$) with a median value of $-0.0038$~\magdsq.
The distribution shows a long tail toward larger negative values.
Most SNe~Ia are observed to have a negative concavity.
A light-curve with a smaller positive \dm\ (\dm$\lesssim0.005$~\magdsq) has a flattened feature, and a light-curve with a larger positive \dm\ (\dm$\gtrsim0.005$~\magdsq) has a similar light-curve shape as a non-detection. 

There are no substantial differences in the \tdm\ and \dm\ distributions between the Gold and Silver samples.
The median values of \tdm\ are 4.3~d and 5.0~d for the Gold and Silver samples, respectively.
The median values of \dm\ are $-0.005$~\magdsq\ and $-0.003$~\magdsq\ for the Gold and Silver samples, respectively.
Compared to the Silver sample, the Gold sample tends to have more SNe~Ia with large negative \dm\ and less SNe~Ia with positive \dm\ (close to non-detections).
This shows the cadence effect discussed in Section~\ref{sec:methods}.
Missing data points at a critical time range can significantly underestimate the strength of the downward concavity.
This effect is included in the larger uncertainties estimated via Monte Carlo iterations (Section~\ref{subsec:measurements}).

\begin{figure}
\centering
\includegraphics[width=\columnwidth]{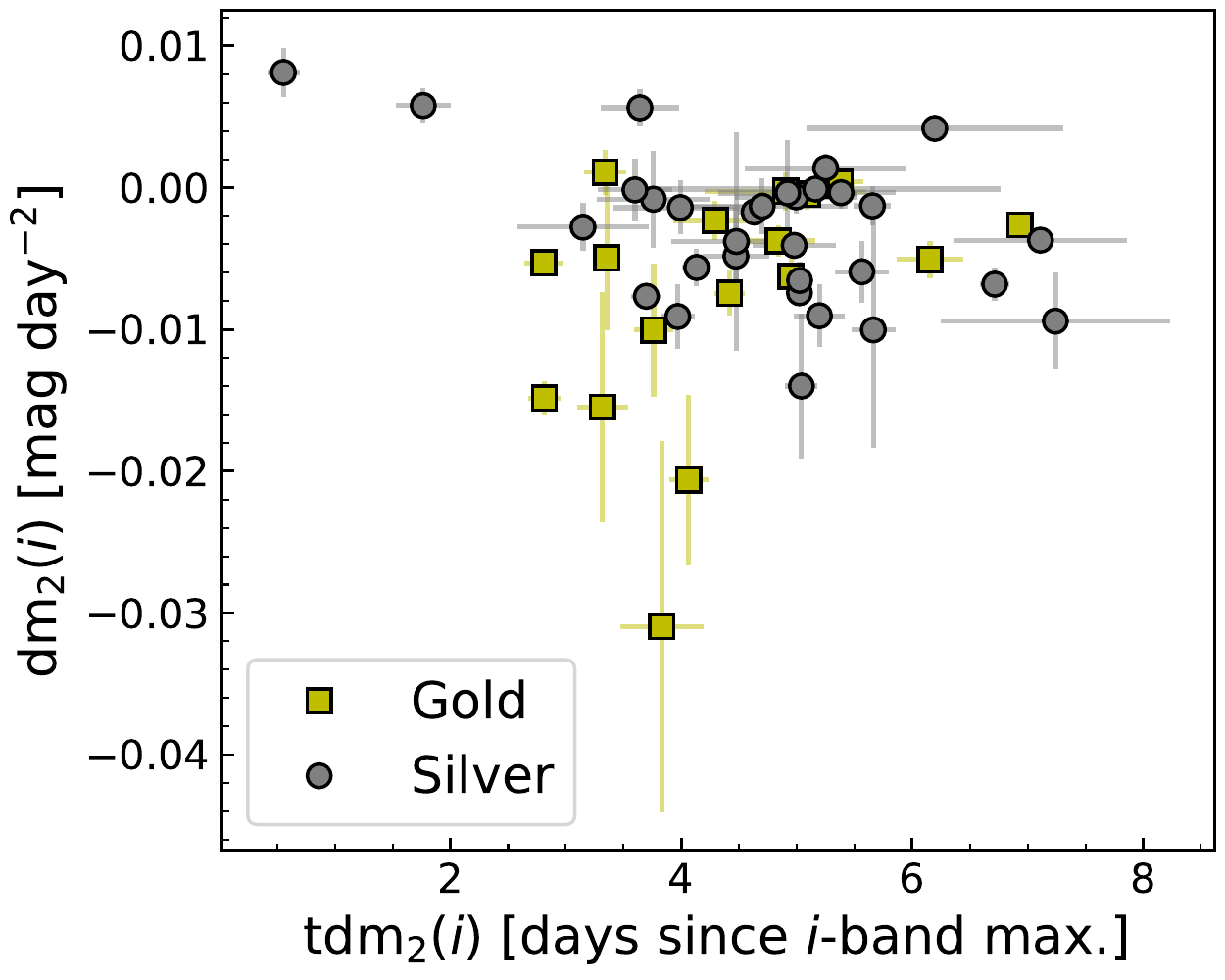}
\caption{
The $i$-band light-curve parameters \tdm\ versus \dm.
The Gold and Silver samples are represented by yellow squares and gray circles, respectively.
}
\label{fig:dmtdm}
\end{figure}

In Figure~\ref{fig:dmtdm}, the \tdm\ and \dm\ parameters are plotted against each other. 
There is no significant correlation between them, suggesting that the two parameters are probing independent properties.
The large negative \dm\ values tend to occur in the narrow \tdm\ range of $\sim3-4$~d but that may be due to the small sample size.

\subsection{Light-curve decline rate}
\label{subsec:sbv}

\begin{figure*}
\centering
\includegraphics[width=\textwidth]{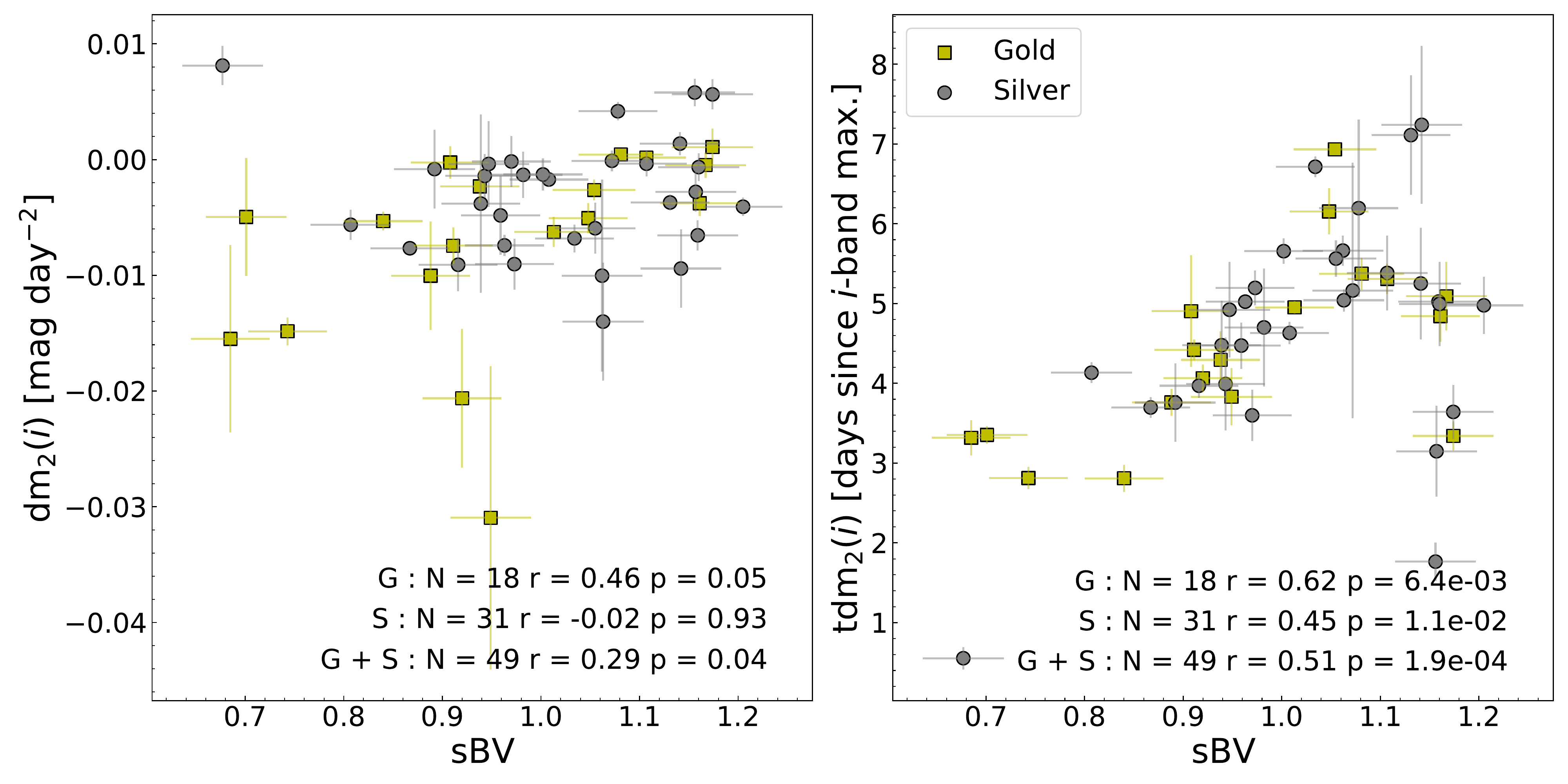}
\caption{
The $i$-band light-curve feature measurements \dm\ (left panel) and \tdm\ (right panel) versus the color-stretch parameter \sBV, a primary SN~Ia parameter.
The Gold and Silver samples are represented by yellow squares and silver circles, respectively.
The number of SNe~Ia ($N$), as well as the Pearson $r$ coefficient and the associated $p$ value are noted for each of the Gold (G), Silver (S), and combined (G + S) samples.
}
\label{fig:sBV}
\end{figure*}

The primary parameter characterizing SNe~Ia is the light-curve decline rate or the color-stretch parameter. 
In order to determine if our \tdm\ and \dm\ measurements provide any new information beyond that provided by the primary parameter, we plot them against each other to search for any correlation in Figure~\ref{fig:sBV}.
We use the color-stretch parameter \sBV\, which was found to provide improved standardization over the classical $B$-band decline-rate measurements, especially in the faint end of the width-luminosity relation \citep{2014ApJ...789...32B}.

The timing of the $i$-band light-curve feature \tdm\ correlates with \sBV\ except for a few outliers at the high-\sBV\ and bright end (right panel of Figure~\ref{fig:sBV}).
The Pearson $r$ coefficient shows a positive correlation ($r=0.51$ for the combined Gold $+$ Silver sample) with the Gold sample displaying a stronger correlation ($r=0.62$).
The $p$-value, denoting the probability of obtaining the current result if the $r$ coefficient is in fact zero, confirms the significance of the correlation with $p=0.00019$ for the combined sample.
The correlation shows that more luminous SNe~Ia with a slower decline rate have a more delayed onset of the $i$-band light-curve feature. At the same time, intrinsically fainter SNe~Ia, ones with more rapidly evolving light curves and lower \sBV, also have an earlier onset of the $i$-band light-curve feature (shorter \tdm).
The \tdm\ parameter therefore does not provide much information independent of nominal $B$ and $V$-band light-curve decline rate measurements.

The strength of the $i$-band light-curve feature \dm, on the other hand, does not show a significant correlation with \sBV.
The Pearson coefficient of $r=0.29$ and the $p$-value of $p=0.04$ for the combined sample are mainly driven by two large negative \dm\ values both landing near \sBV$=0.9$. 
A Bayesian approach was then considered to test this result by including the errors associated to the measurements. Using the {\sc LINMIX} package, which implements the Hierarchical Bayesian model of \citet{2007ApJ...665.1489K}, we again see only a weak ($\sim$~2$\sigma$) trend.
Therefore, we conclude that the \dm\ parameter is largely independent of \sBV.
Given the independent nature of the \dm\ measurement, we explore its relation with other SN~Ia properties in detail in the following subsections.

Other properties at maximum light were also examined.
The reddening corrected $B-V$, $r-i$, and $B-i$ pseudo-colors\footnote{Colors calculated as the differences of respective peak magnitudes in each band.} were determined using the peak magnitude of each respective band via the GP interpolation. 
No significant correlations were found with either \tdm\ or \dm, with the most significant Pearson coefficient of $r=-0.46$ identified for $B-i$ versus \tdm.
The timing of the primary $i$-band maximum relative to that of $B$-band is also found to have no significant correlation with either \tdm\ or \dm.

\subsection{Hubble residuals}
\label{subsec:res}

One of the key objectives of this project is to determine whether the $i$-band light-curves of SNe~Ia provide information independent of the $B$ and $V$ bands and whether that information could provide improved standardization of the SN~Ia peak luminosity.

\begin{figure*}
\centering
\includegraphics[width=\textwidth]{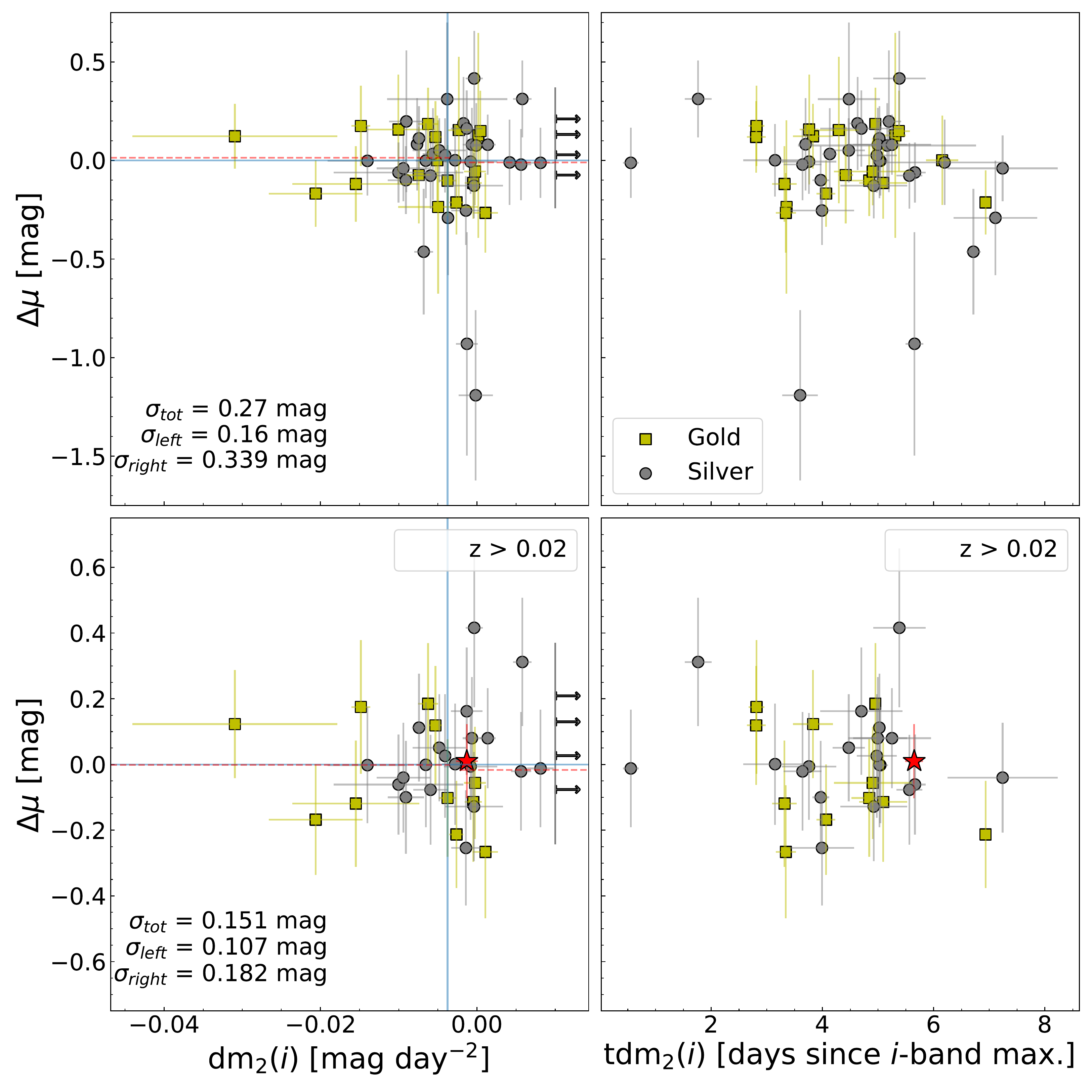}
\caption{The $i$-band light-curve parameters \dm\ (left panels) and \tdm\ (right panels) versus the Hubble residuals.
The full sample is plotted in the top panels, while in the bottom panels, only SNe~Ia at z$_{cmb} > 0.02$ plus the special case of SN~2013aa (red star), are included to minimize the distance uncertainty from the host galaxy peculiar velocities.
On the left panels, the blue vertical lines show the sample \dm\ median values, and the red horizontal dashed lines show the mean of the Hubble residuals at each side of \dm\ median. 
The Hubble diagram scatters for the whole sample ($\sigma_\mathrm{tot}$), and for SNe~Ia with \dm\ less than ($\sigma_\mathrm{left}$) and greater ($\sigma_\mathrm{right}$) than the median value are also noted. 
The Gold and Silver samples are represented by yellow squares and silver circles, respectively. Silver arrows show the Hubble residuals of the SNe in the silver sample for which a feature was not detected.
}
\label{fig:residuals}
\end{figure*}

The $i$-band light-curve parameters \tdm\ and \dm\ are plotted versus the Hubble residuals in Figure~\ref{fig:residuals}.
The Hubble residuals were computed as the difference between the distance moduli determined from the SNe~Ia themselves using the methods outlined in \citet{2018ApJ...869...56B} and that determined using the host galaxy redshifts corrected to the rest frame of the cosmic microwave background and the assumed cosmology: $H_0=72$ km s$^{-1}$ Mpc$^{-1}$, $\Omega_M=0.27$, and $\Omega_{\Lambda}=0.73$.
The uncertainties for the Hubble residuals include SN~Ia observational and light-curve fitting uncertainties, as well as that from the host galaxy peculiar velocity which is assumed to be 350~\kms.

To minimize the effect of peculiar velocities, we also present the same figure with only SNe~Ia at z$_{cmb} > 0.02$, where peculiar velocities are much smaller than the Universal expansion (bottom panels of Figure~\ref{fig:residuals}).
SN~2013aa (presented as a red star in Figure~\ref{fig:residuals}) is a special case in that it is a sibling to SN~2017cbv (they have the same galaxy host, NGC~5643). In this case, we can compute a residual independent of the Hubble law and hence avoid the large uncertainty due to its low redshift. We use the Phillips relation to estimate the distance to SN~2017cbv and use this as a distance to NGC~5643, which we can then compare with the distance from SN~2013aa \citep{2020ApJ...895..118B}. Using this procedure, we derive a residual of -0.01 mag +/- 0.14 mag. While this may seem circular, we are only interested in the {\em relative} distances predicted by the two SNe~Ia, so any systematic errors in the Phillips relation have very little effect on the result. It is interesting that these two SNe~Ia have very similar \dm\ values.

The timing parameter \tdm\ versus Hubble residual plot does not show any interesting features.
This is expected as \tdm\ strongly correlates with \sBV.
On the other hand, the \dm\ versus Hubble residual plot shows a smaller scatter for SNe~Ia with more negative \dm.
This effect is obvious when the redshift cut is imposed and the uncertainty from peculiar velocities is minimized.

In effect, the \dm\ parameter is able to isolate a less scattered SN~Ia sample.
We do not yet understand the physical mechanism for this effect, but the cosmological potential utility is obvious.
Our combined Gold and Silver sample with z$_{cmb} > 0.02$ yields a scatter on the Hubble diagram of 0.151~mag.
If the sample is divided by the median \dm\ at $-0.0038$ \magdsq, SNe~Ia with more negative \dm\ values (stronger $i$-band light-curve features, 15 SNe) yield a scatter on the Hubble diagram of $\sigma_\mathrm{left} = 0.107$ mag, compared to $\sigma_\mathrm{right} = 0.182$ mag for the rest of the sample (16 SNe).
By selecting only SNe~Ia with strong $i$-band features of downward concavity, the improvement in distance accuracy is substantial (for this small sample), akin to, for example, shifting observations to the NIR \citep[e.g.,][]{2012MNRAS.425.1007B, 2019ApJ...887..106A}, standardization using spectral features \citep[e.g.,][]{2009A&A...500L..17B} or avoiding the centers of the host galaxies \citep[e.g.,][]{2012ApJ...755..125G, 2020ApJ...901..143U}.

The bootstrap method was used to determine whether the measured scatter in the Hubble diagram is due to small-number statistics.
We consider 10{,}000 bootstrap re-samples, divide the results by the median of each re-sample, and calculate the Hubble diagram dispersion on each side of the median. We see that only $\sim$ 6$\%$ of the re-samples show $\sigma _\mathrm{right} \leq \sigma_\mathrm{left}$, with a mean difference between $\sigma _\mathrm{right}$ and $\sigma_\mathrm{left}$ of 0.009. The remaining 94$\%$ of the re-samples show $\sigma _\mathrm{right} \geq \sigma_\mathrm{left}$ with mean values of $\mu_{\sigma _\mathrm{right}} = 0.17 \pm 0.04$ and $\mu_{\sigma _\mathrm{left}} = 0.1 \pm 0.01$, consistent with the original results.
A Kolmogorov-Smirnov test was also performed resulting in a statistic value of 0.24. Hence, while the differences observed in the dispersion of the residuals are intriguing, the result is not currently statistically significant. Larger samples should be obtained in the future to further test whether selecting SNe~Ia by their \dm\ can reduce Hubble residuals.

\subsection{Environmental properties}
\label{subsec:host}

SN~Ia environment and host galaxy properties are known to be linked to SN~Ia properties \citep[e.g.,][]{2000AJ....120.1479H} and are crucial in deciphering the origins of these explosions.
They have also been shown to affect their distance determination.
Evidence has been presented that after standardization using their light-curve shapes and colors, SNe~Ia are on average more luminous in massive host galaxies \citep[e.g.,][]{2010MNRAS.406..782S}.
Furthermore, SNe~Ia at larger projected host galactocentric distances show a smaller scatter on the Hubble diagram \citep{2012ApJ...755..125G, 2020ApJ...901..143U} with improvements similar to what we have shown in Section~\ref{subsec:res}.
To check if these environmental effects are linked to our findings, we examine the relationship between the $i$-band light-curve parameters and the environmental properties.
All environmental properties shown in this work were adopted from \citet{2020ApJ...901..143U}.

\begin{figure*}[t]
\centering
\includegraphics[width=\textwidth]{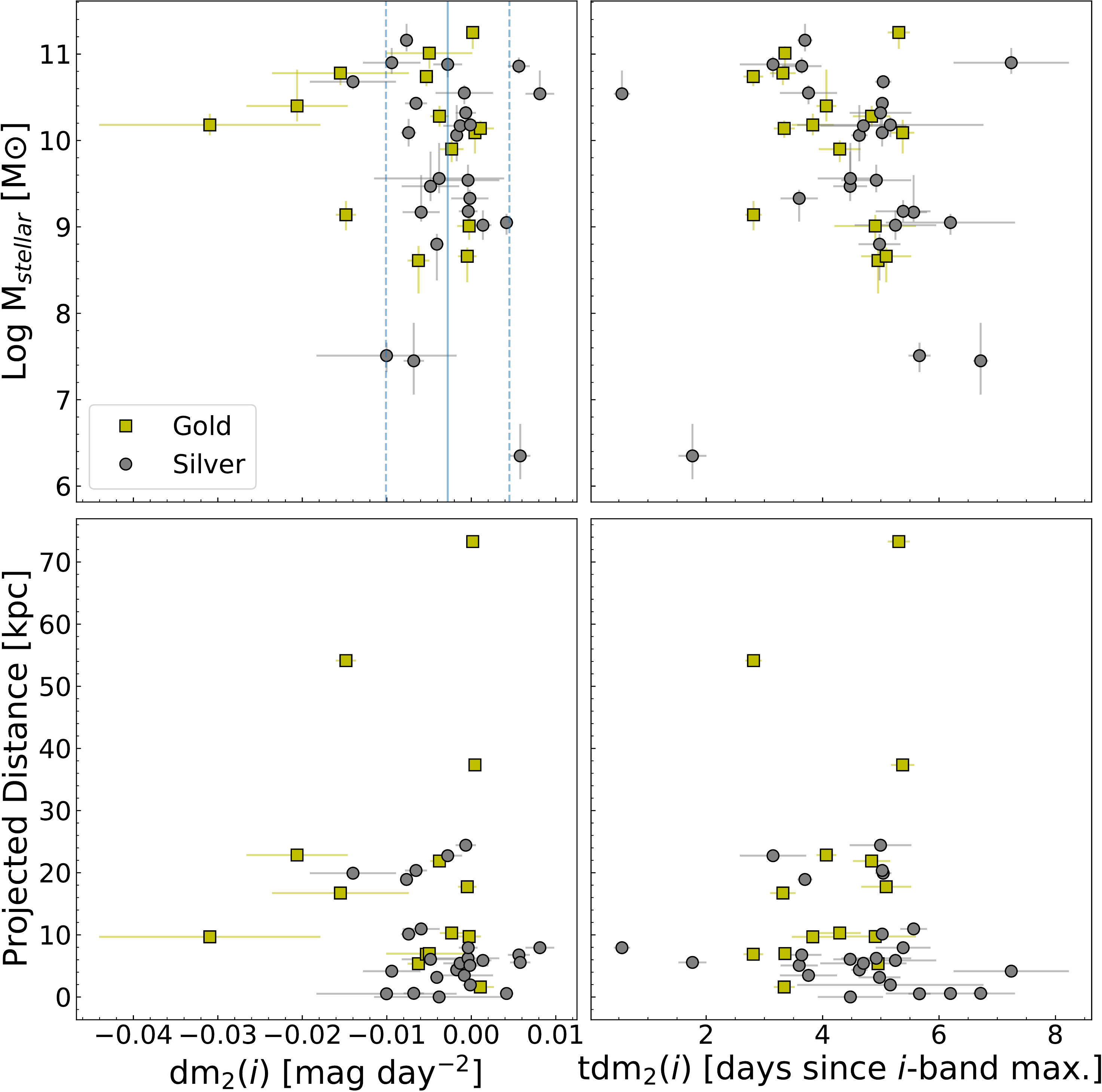}
\caption{
The $i$-band light-curve parameters \dm\ (left panels) and \tdm\ (right panels) versus host galaxy stellar mass (top panels) and projected host galactocentric distance (bottom panels).
For the top left panel, the solid blue vertical line locate the median \dm\ of the entire sample, and the dashed blue lines represent the 1-$\sigma$ dispersion from the median.
The Gold and Silver samples are represented by yellow squares and silver circles, respectively.
}
\label{fig:mass}
\end{figure*}

There are no significant correlations found between the \dm\ and \tdm\ parameters and the host galaxy stellar mass (top panels of Figure~\ref{fig:mass}).
It is worth noting that SNe~Ia with very strong $i$-band features (large negative \dm) come exclusively from host galaxies with high stellar mass. 
However, the sample of these extreme light-curves is small and they have similar \sBV\ values.
The top left panel of Figure~\ref{fig:mass} illustrates the effect with all SNe~Ia with \dm\ more than 1-$\sigma$ below the sample median coming from host galaxies with stellar mass greater than $10^9$~M$_{\odot}$.
On the other hand, SNe~Ia with moderate or no $i$-band features come from host galaxies with the full range of stellar mass.

There are also no significant correlations found between the \dm\ and \tdm\ parameters and the projected host galactocentric distance of the SNe~Ia (bottom panels of Figure~\ref{fig:mass}).
SNe~Ia at large projected host galactocentric distances (defined as greater than 10~kpc in  \citealt{2020ApJ...901..143U}), on average, were shown to have smaller Hubble residuals.
In Section~\ref{subsec:res}, SNe~Ia with more negative \dm, on average, were shown to smaller Hubble residuals.
The bottom left panel of Figure~\ref{fig:mass} confirms that the above are two independent effects.
SNe~Ia that explode far from the host galaxy centers have a wide range of strengths in their $i$-band light-curve feature.

\subsection{Spectroscopic properties}
\label{subsec:spec}

The $i$-band spectral region is dominated by P Cygni features formed by intermediate-mass elements, such as \ion{O}{I}, \ion{Mg}{II}, and \ion{Ca}{II}.
In particular, the \ion{Ca}{II} infrared triplet forms a strong absorption feature that dominates the region.
In many early-phase spectra, high-velocity \ion{Ca}{II} features are observed to be detached from the photospheric component and can persist past maximum light in some SNe~Ia \citep[e.g.,][]{2005ApJ...623L..37M}.
The spectral profile shape and strength in the $i$-band are largely dictated by the light-curve decline rate or intrinsic brightness of SNe~Ia \citep[e.g.,][]{2009PhDT.......228H}.

We have observed in Section~\ref{subsec:res} an interesting difference between two samples divided by the median \dm\ value which measures the strength of the $i$-band light-curve feature.
SNe~Ia with a stronger feature tend to be more uniform in their distance determination after standardization, as shown by their smaller Hubble diagram scatter.
If the difference is linked to a physical effect, such as the speed of the ionization evolution \citep[e.g.,][]{2006ApJ...649..939K}, it may be reflected in the spectral features.

\begin{figure}
\centering
\includegraphics[width=\columnwidth]{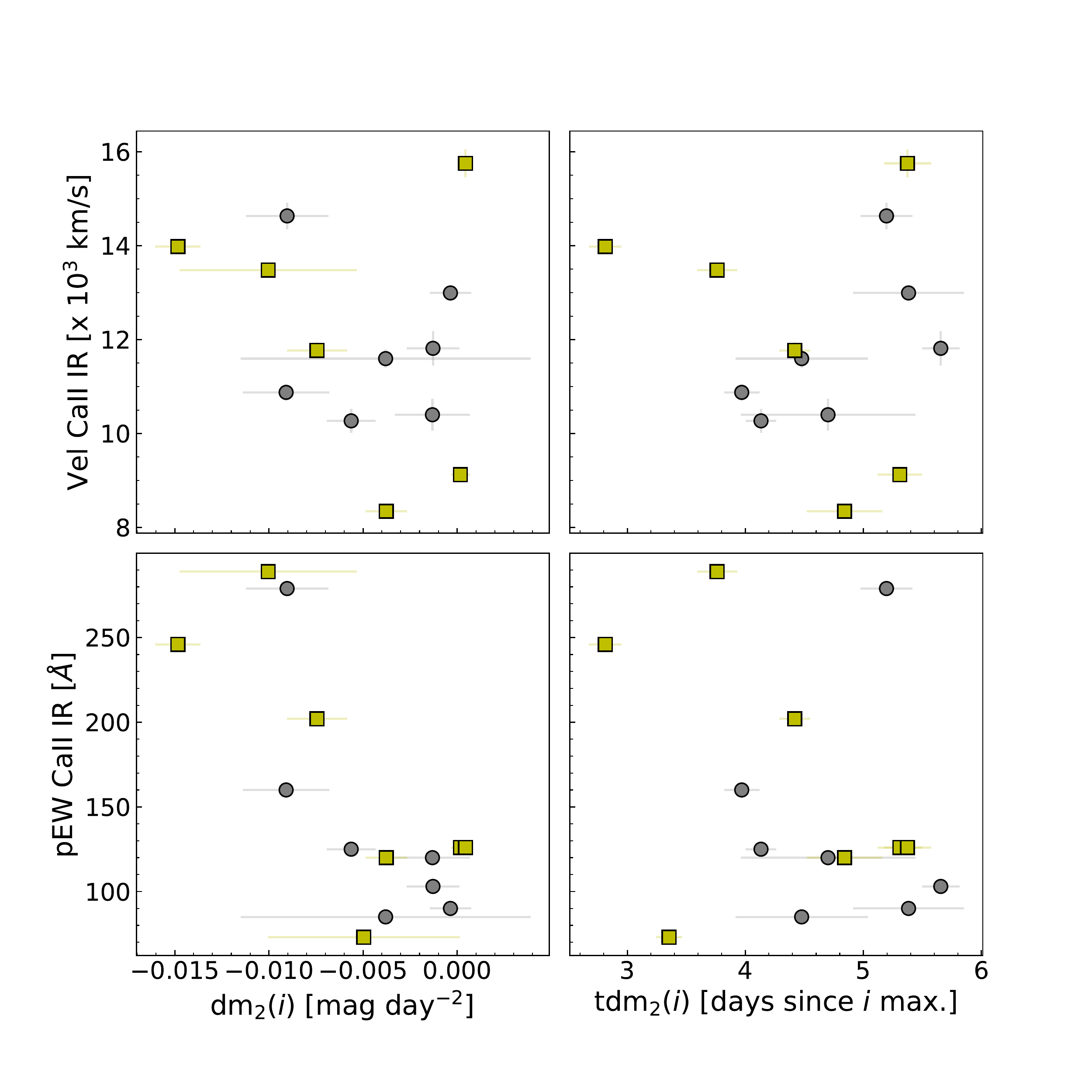}
\caption{
The $i$-band light-curve parameters \dm\ (left panels) and \tdm\ (right panels) versus \ion{Ca}{II} infrared triplet velocity (top panels) and \ion{Ca}{II} infrared triplet pEW (bottom panels) at $B$-band maximum.
The Gold and Silver samples are represented by yellow squares and silver circles, respectively.
}
\label{fig:caIIIR}
\end{figure}

We specifically examine the velocity and the pseudo equivalent width (pEW) of the \ion{Ca}{II} infrared triplet at $B$-band maximum, which is the strongest spectral feature in the $i$-band (Figure~\ref{fig:caIIIR}).
The \dm\ parameter shows a strong correlation with the pEW of the \ion{Ca}{II} infrared triplet (Pearson coefficient $r=-0.76$ and a p value of $0.001$).
SN~Ia with a strong light-curve feature with a downward concavity also have a stronger \ion{Ca}{II} spectral feature.
No other significant correlations were found between the $i$-band light-curve parameters and the properties of the \ion{Ca}{II} infrared triplet.

\begin{figure}
\centering
\includegraphics[width=\columnwidth]{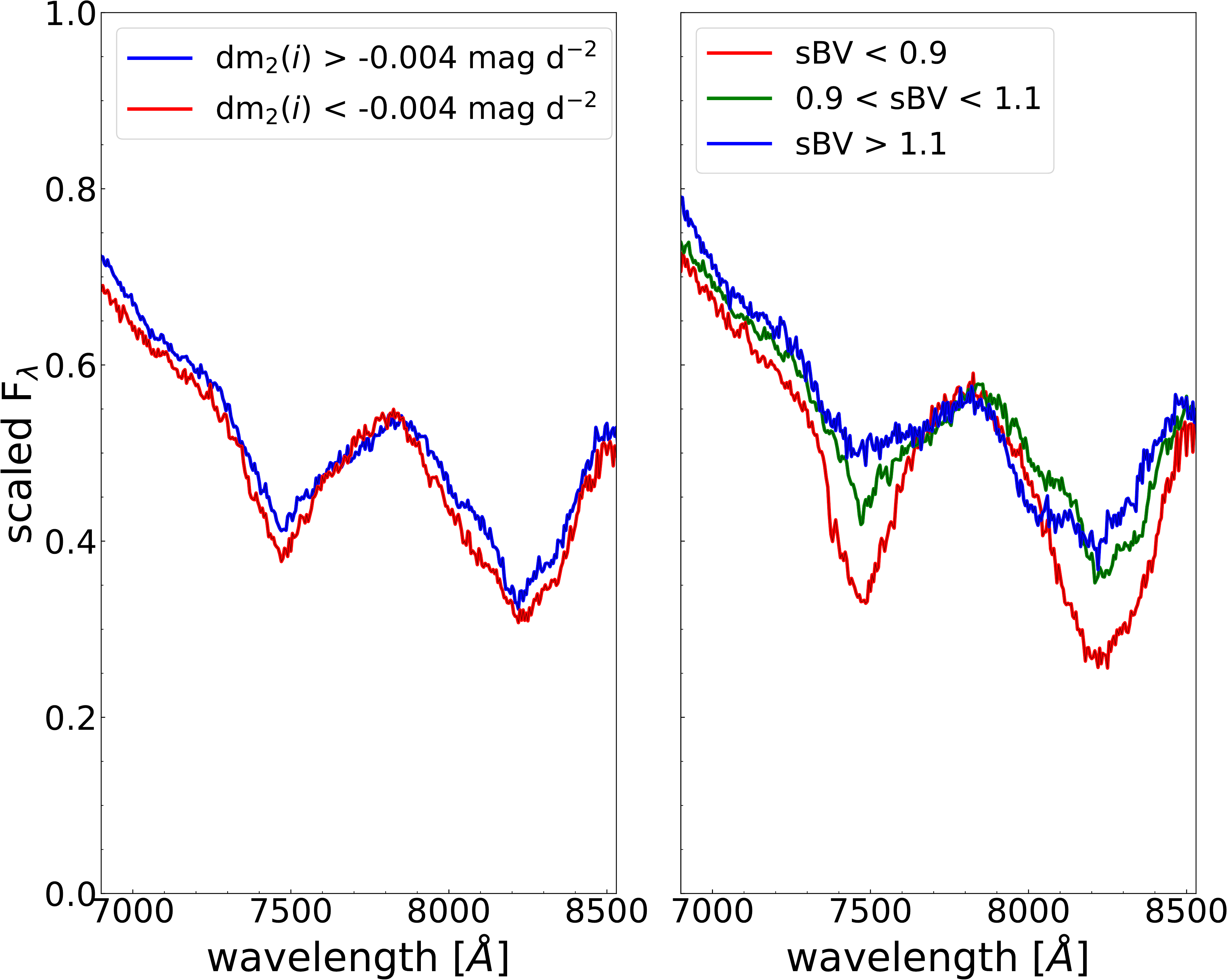}
\caption{
The mean spectrum of bootstrapped samples comparing the $i$-band spectral features of SNe~Ia of various \dm\ and \sBV. 
For each SN~Ia, the spectrum observed closest to \tdm\ is included in the sample with a $\pm5$~d window. 
The left panel shows the mean spectra from two groups separated by \dm$=-0.004$ \magdsq.
For comparison, the right panel shows the mean spectra from three groups separated by their \sBV.
}
\label{fig:spec}
\end{figure}

In an attempt to detect subtle differences in the $i$-band spectral profiles between the high and low-\dm\ groups, we examine the mean spectrum of the bootstrapped sample of each group.
The two groups are divided by the median \dm\ of the entire sample.
For each SN~Ia, we include the spectrum observed closest to \tdm\ within a $\pm5$~d window.
Only one spectrum per SN~Ia is included to prevent a few well-observed SNe~Ia from dominating the results.
The bootstrapped mean spectrum for each \dm\ group is presented in the left panel of Figure~\ref{fig:spec}.

There are no substantial spectral differences found between the two samples separated by \dm.
Specifically, the two bootstrapped mean spectra show the same features with only slightly different absorption strengths.
The correlation between the \dm\ parameter and the pEW of \ion{Ca}{II} infrared triplet is confirmed.
The group of SNe~Ia with more negative \dm\ values, i.e., ones with the stronger light-curve features and smaller Hubble diagram scatter, have stronger \ion{Ca}{II} infrared triplet and \ion{O}{I} absorptions on average.
The radial velocity shifts of these absorptions are also similar in the two groups.

Note that these differences between the two \dm-divided groups are small compared to those driven by \sBV\ known previously.
The right panel of Figure~\ref{fig:spec} shows the bootstrapped mean spectra of the same SNe~Ia but now divided by \sBV.
Much clearer differences are seen, both in the absorption strength and profile shapes.
The high-\sBV\ (intrinsically bright) SNe~Ia have a weaker \ion{Ca}{II} infrared triplet and \ion{O}{I} absorption and evidence of a persistent high-velocity \ion{Ca}{II} component when compared to the low-\sBV\ objects.

\section{Conclusions and discussion}
\label{sec:conclusion}

In this work, we have investigated a peculiar feature in the $i$-band light-curves present in many SNe~Ia.
The feature emerges between 0.5 and 7.5~d relative to $i$-band maximum in our sample, well before the secondary maximum.
It is an abrupt change in curvature in the light-curve over a few days and appears as a flattening in the mild cases and a strong downward concave shape or a ``kink'' in the most extreme cases. We do not see such a feature systematically present in bluer bands.

Since the feature can be subtle at times and lasts only for a few days, a data set of high S/N ratio and rapid cadence is required.
The light-curves obtained by the CSP are ideal for this study.
A sample of $i$-band light-curves was compiled from 54 nearby SNe~Ia, of which, 19 have the optimal nightly cadence in the phase range of interest (Gold sample) and 35 have a slightly lower maximum cadence of up to 1.5 rest-frame days (Silver sample).
To quantify the light-curve feature, the observed data points were interpolated using GP, and the second derivative of the interpolated light-curve was computed for each light-curve to detect any change in curvature.

Two parameters were measured to characterize the $i$-band light-curve feature. 
The \tdm\ parameter is measured in days and specifies the timing of the feature relative to $i$-band maximum.
The \dm\ parameter is measured in \magdsq\ and represents the strength and direction of the concavity.
In our combined Gold and Silver sample, 76\% of the SNe~Ia show a negative \dm, representing a downward concavity, either a mild flattening or a strong ``kink.''
The rest either have a small positive \dm\ (still a weak flattening) or are deemed non-detections.
There appears to be a continuum of \dm\ values from very negative to slightly positive values.

SNe~Ia are remarkably uniform, and their various observed properties are largely dictated by the primary parameter, such as the $B$-band decline rate or \sBV.
It is therefore invaluable to find an independent secondary or tertiary parameter to provide more information on SNe~Ia and possibly improve their standardization.
The \tdm\ parameter is shown to correlate with the light-curve parameter \sBV, as the timing of the $i$-band feature is largely affected by the time-axis stretching of the $i$-band light-curve.
Therefore, the \tdm\ parameter does not provide extra information beyond that given by a primary parameter.
On the other hand, the \dm\ parameter does not correlate strongly with \sBV\ and is deemed an independent parameter.
It is also largely independent of the information provided by spectroscopic and host galaxy properties.

A potentially very interesting result of this work comes from the examination of Hubble residuals from the SNe~Ia in our sample.
The SNe~Ia with large negative \dm\ values (ones with the strongest downward concave feature) also show the smallest Hubble residuals.
The effect is more apparent when the sample is limited to SNe~Ia in the Hubble flow (z$_{cmb} > 0.02$) where the uncertainty from the host galaxy peculiar velocities is minimized.
If we divide the Hubble-flow sample in half along its median \dm\ value, the sample with more negative \dm\ values has a Hubble diagram dispersion of 0.107~mag, 0.075~mag smaller than the rest.
This result may offer a new method for improving SNe~Ia as standard candles without shifting to more costly NIR or spectroscopic observations, however larger statistics are required for stronger conclusions.

We do not yet understand the physical processes that form the observed $i$-band light-curve feature and why SNe~Ia with stronger features have smaller scatter on the Hubble diagram. 
Is the feature formed by added flux perhaps from the light echoes reflected off circumstellar material \citep{2015MNRAS.452.3281M}?
If so, the peculiarity in the light-curve would be even stronger in the NIR.
We currently do not have adequate NIR light-curves to check this.
It is also not intuitive that SNe~Ia with the stronger features and presumably stronger light echoes would be more uniform in their distance determinations.
A well-known example of a SN~Ia with light echoes is SN~2006X \citep{2008ApJ...677.1060W}.
This SN is in our sample and shows only a weak $i$-band light-curve flattening (\dm$=-0.0002$ \magdsq). 
Is the light-curve feature formed by an ionization effect, e.g., from \ion{Fe}{III} to \ion{Fe}{II}?
Or is it formed by Ca plumes arising from instabilities in the deflagration front \citep[e.g.,][]{1995ApJ...449..695K}?
For both of these scenarios, the effect should be detected in spectral features. We do see a strong correlation between the \dm\ parameter and the pEW of the \ion{Ca}{II} infrared triplet which is strongly temperature sensitive but the bootstrapped analysis of the spectral lines show almost no differences.
However, these possibilities are purely speculative at this point and further theoretical modeling work is needed to explain these observations.

In this work we have robustly identified a previously overlooked feature in SN~Ia light curves, and characterised its properties. Future investigation should attempt to further understand its physical origin, and test the significance of differences in Hubble residuals observed in our work.

\section*{Acknowledgements}

We thank the referee for his comments and for carefully reading the manuscript. The work of the CSP has been supported by the National Science Foundation under grants AST0306969, AST0607438, AST1008343, AST1613426, AST1613455 and AST1613472. P.J.P thanks the kindness and support of the FSU SN team. L.G. acknowledges financial support from the Spanish Ministry of Science and Innovation (MCIN) under the 2019 Ram\'on y Cajal program RYC2019-027683 and from the Spanish MCIN project HOSTFLOWS PID2020-115253GA-I00. This research has made use of the NASA/IPAC Extragalactic Database (NED), which is operated by the Jet Propulsion Laboratory, California Institute of Technology, under contract with the National Aeronautics and Space Administration.

\section*{Data Availability}

The data used on this paper will be shared upon reasonable request. 

\section*{software}
{\sc numpy} \citep{harris2020array},
{\sc matplotlib} \citep{Hunter:2007}, 
{\sc pandas} \citep{mckinney-proc-scipy-2010},
{\sc scipy} \citep{2020SciPy-NMeth},
{\sc astropy} \citep{astropy:2018},
{\sc LINMIX},
{\sc GPy} \citep{gpy2014}



\bibliographystyle{mnras}
\bibliography{SNIa_kink} 

\begin{thebibliography}{}
\makeatletter
\relax
\def\mn@urlcharsother{\let\do\@makeother \do\$\do\&\do\#\do\^\do\_\do\%\do\~}
\def\mn@doi{\begingroup\mn@urlcharsother \@ifnextchar [ {\mn@doi@}
  {\mn@doi@[]}}
\def\mn@doi@[#1]#2{\def\@tempa{#1}\ifx\@tempa\@empty \href
  {http://dx.doi.org/#2} {doi:#2}\else \href {http://dx.doi.org/#2} {#1}\fi
  \endgroup}
\def\mn@eprint#1#2{\mn@eprint@#1:#2::\@nil}
\def\mn@eprint@arXiv#1{\href {http://arxiv.org/abs/#1} {{\tt arXiv:#1}}}
\def\mn@eprint@dblp#1{\href {http://dblp.uni-trier.de/rec/bibtex/#1.xml}
  {dblp:#1}}
\def\mn@eprint@#1:#2:#3:#4\@nil{\def\@tempa {#1}\def\@tempb {#2}\def\@tempc
  {#3}\ifx \@tempc \@empty \let \@tempc \@tempb \let \@tempb \@tempa \fi \ifx
  \@tempb \@empty \def\@tempb {arXiv}\fi \@ifundefined
  {mn@eprint@\@tempb}{\@tempb:\@tempc}{\expandafter \expandafter \csname
  mn@eprint@\@tempb\endcsname \expandafter{\@tempc}}}

\bibitem[\protect\citeauthoryear{{Ashall} et~al.,}{{Ashall}
  et~al.}{2020}]{2020ApJ...895L...3A}
{Ashall} C.,  et~al., 2020, \mn@doi [\apjl] {10.3847/2041-8213/ab8e37}, \href
  {https://ui.adsabs.harvard.edu/abs/2020ApJ...895L...3A} {895, L3}

\bibitem[\protect\citeauthoryear{{Astropy Collaboration} et~al.,}{{Astropy
  Collaboration} et~al.}{2018}]{astropy:2018}
{Astropy Collaboration} et~al., 2018, \mn@doi [\aj] {10.3847/1538-3881/aabc4f},
  \href {https://ui.adsabs.harvard.edu/abs/2018AJ....156..123A} {156, 123}

\bibitem[\protect\citeauthoryear{{Avelino}, {Friedman}, {Mandel}, {Jones},
  {Challis}  \& {Kirshner}}{{Avelino} et~al.}{2019}]{2019ApJ...887..106A}
{Avelino} A.,  {Friedman} A.~S.,  {Mandel} K.~S.,  {Jones} D.~O.,  {Challis}
  P.~J.,   {Kirshner} R.~P.,  2019, \mn@doi [\apj] {10.3847/1538-4357/ab2a16},
  \href {https://ui.adsabs.harvard.edu/abs/2019ApJ...887..106A} {887, 106}

\bibitem[\protect\citeauthoryear{{Bailey} et~al.,}{{Bailey}
  et~al.}{2009}]{2009A&A...500L..17B}
{Bailey} S.,  et~al., 2009, \mn@doi [\aap] {10.1051/0004-6361/200911973}, \href
  {https://ui.adsabs.harvard.edu/abs/2009A&A...500L..17B} {500, L17}

\bibitem[\protect\citeauthoryear{{Barone-Nugent} et~al.,}{{Barone-Nugent}
  et~al.}{2012}]{2012MNRAS.425.1007B}
{Barone-Nugent} R.~L.,  et~al., 2012, \mn@doi [\mnras]
  {10.1111/j.1365-2966.2012.21412.x}, \href
  {https://ui.adsabs.harvard.edu/abs/2012MNRAS.425.1007B} {425, 1007}

\bibitem[\protect\citeauthoryear{{Burns} et~al.,}{{Burns}
  et~al.}{2011}]{2011AJ....141...19B}
{Burns} C.~R.,  et~al., 2011, \mn@doi [\aj] {10.1088/0004-6256/141/1/19}, \href
  {https://ui.adsabs.harvard.edu/abs/2011AJ....141...19B} {141, 19}

\bibitem[\protect\citeauthoryear{{Burns} et~al.,}{{Burns}
  et~al.}{2014}]{2014ApJ...789...32B}
{Burns} C.~R.,  et~al., 2014, \mn@doi [\apj] {10.1088/0004-637X/789/1/32},
  \href {https://ui.adsabs.harvard.edu/abs/2014ApJ...789...32B} {789, 32}

\bibitem[\protect\citeauthoryear{{Burns} et~al.,}{{Burns}
  et~al.}{2018}]{2018ApJ...869...56B}
{Burns} C.~R.,  et~al., 2018, \mn@doi [\apj] {10.3847/1538-4357/aae51c}, \href
  {https://ui.adsabs.harvard.edu/abs/2018ApJ...869...56B} {869, 56}

\bibitem[\protect\citeauthoryear{{Burns} et~al.,}{{Burns}
  et~al.}{2020}]{2020ApJ...895..118B}
{Burns} C.~R.,  et~al., 2020, \mn@doi [\apj] {10.3847/1538-4357/ab8e3e}, \href
  {https://ui.adsabs.harvard.edu/abs/2020ApJ...895..118B} {895, 118}

\bibitem[\protect\citeauthoryear{{Childress}, {Filippenko}, {Ganeshalingam}  \&
  {Schmidt}}{{Childress} et~al.}{2014}]{2014MNRAS.437..338C}
{Childress} M.~J.,  {Filippenko} A.~V.,  {Ganeshalingam} M.,   {Schmidt} B.~P.,
   2014, \mn@doi [\mnras] {10.1093/mnras/stt1892}, \href
  {https://ui.adsabs.harvard.edu/abs/2014MNRAS.437..338C} {437, 338}

\bibitem[\protect\citeauthoryear{{Contreras} et~al.,}{{Contreras}
  et~al.}{2010}]{2010AJ....139..519C}
{Contreras} C.,  et~al., 2010, \mn@doi [\aj] {10.1088/0004-6256/139/2/519},
  \href {https://ui.adsabs.harvard.edu/abs/2010AJ....139..519C} {139, 519}

\bibitem[\protect\citeauthoryear{{Fakhouri} et~al.,}{{Fakhouri}
  et~al.}{2015}]{2015ApJ...815...58F}
{Fakhouri} H.~K.,  et~al., 2015, \mn@doi [\apj] {10.1088/0004-637X/815/1/58},
  \href {https://ui.adsabs.harvard.edu/abs/2015ApJ...815...58F} {815, 58}

\bibitem[\protect\citeauthoryear{{Folatelli} et~al.,}{{Folatelli}
  et~al.}{2010}]{2010AJ....139..120F}
{Folatelli} G.,  et~al., 2010, \mn@doi [\aj] {10.1088/0004-6256/139/1/120},
  \href {https://ui.adsabs.harvard.edu/abs/2010AJ....139..120F} {139, 120}

\bibitem[\protect\citeauthoryear{{Foley}, {Narayan}, {Challis}, {Filippenko},
  {Kirshner}, {Silverman}  \& {Steele}}{{Foley}
  et~al.}{2010}]{2010ApJ...708.1748F}
{Foley} R.~J.,  {Narayan} G.,  {Challis} P.~J.,  {Filippenko} A.~V.,
  {Kirshner} R.~P.,  {Silverman} J.~M.,   {Steele} T.~N.,  2010, \mn@doi [\apj]
  {10.1088/0004-637X/708/2/1748}, \href
  {https://ui.adsabs.harvard.edu/abs/2010ApJ...708.1748F} {708, 1748}

\bibitem[\protect\citeauthoryear{{Freedman} et~al.,}{{Freedman}
  et~al.}{2009}]{2009ApJ...704.1036F}
{Freedman} W.~L.,  et~al., 2009, \mn@doi [\apj] {10.1088/0004-637X/704/2/1036},
  \href {https://ui.adsabs.harvard.edu/abs/2009ApJ...704.1036F} {704, 1036}

\bibitem[\protect\citeauthoryear{{GPy}}{{GPy}}{2012}]{gpy2014}
{GPy} since 2012, {GPy}: A Gaussian process framework in python,
  \url{http://github.com/SheffieldML/GPy}

\bibitem[\protect\citeauthoryear{{Galbany} et~al.,}{{Galbany}
  et~al.}{2012}]{2012ApJ...755..125G}
{Galbany} L.,  et~al., 2012, \mn@doi [\apj] {10.1088/0004-637X/755/2/125},
  \href {https://ui.adsabs.harvard.edu/abs/2012ApJ...755..125G} {755, 125}

\bibitem[\protect\citeauthoryear{{Gall} et~al.,}{{Gall}
  et~al.}{2018}]{2018A&A...611A..58G}
{Gall} C.,  et~al., 2018, \mn@doi [\aap] {10.1051/0004-6361/201730886}, \href
  {https://ui.adsabs.harvard.edu/abs/2018A&A...611A..58G} {611, A58}

\bibitem[\protect\citeauthoryear{{Gonz{\'a}lez-Gait{\'a}n}
  et~al.,}{{Gonz{\'a}lez-Gait{\'a}n} et~al.}{2014}]{2014ApJ...795..142G}
{Gonz{\'a}lez-Gait{\'a}n} S.,  et~al., 2014, \mn@doi [\apj]
  {10.1088/0004-637X/795/2/142}, \href
  {https://ui.adsabs.harvard.edu/abs/2014ApJ...795..142G} {795, 142}

\bibitem[\protect\citeauthoryear{{Hamuy}, {Phillips}, {Suntzeff}, {Schommer},
  {Maza}, {Smith}, {Lira}  \& {Aviles}}{{Hamuy}
  et~al.}{1996}]{1996AJ....112.2438H}
{Hamuy} M.,  {Phillips} M.~M.,  {Suntzeff} N.~B.,  {Schommer} R.~A.,  {Maza}
  J.,  {Smith} R.~C.,  {Lira} P.,   {Aviles} R.,  1996, \mn@doi [\aj]
  {10.1086/118193}, \href
  {https://ui.adsabs.harvard.edu/abs/1996AJ....112.2438H} {112, 2438}

\bibitem[\protect\citeauthoryear{{Hamuy}, {Trager}, {Pinto}, {Phillips},
  {Schommer}, {Ivanov}  \& {Suntzeff}}{{Hamuy}
  et~al.}{2000}]{2000AJ....120.1479H}
{Hamuy} M.,  {Trager} S.~C.,  {Pinto} P.~A.,  {Phillips} M.~M.,  {Schommer}
  R.~A.,  {Ivanov} V.,   {Suntzeff} N.~B.,  2000, \mn@doi [\aj]
  {10.1086/301527}, \href
  {https://ui.adsabs.harvard.edu/abs/2000AJ....120.1479H} {120, 1479}

\bibitem[\protect\citeauthoryear{{Hamuy} et~al.,}{{Hamuy}
  et~al.}{2003}]{2003Natur.424..651H}
{Hamuy} M.,  et~al., 2003, \mn@doi [\nat] {10.1038/nature01854}, \href
  {https://ui.adsabs.harvard.edu/abs/2003Natur.424..651H} {424, 651}

\bibitem[\protect\citeauthoryear{{Hamuy} et~al.,}{{Hamuy}
  et~al.}{2006}]{2006PASP..118....2H}
{Hamuy} M.,  et~al., 2006, \mn@doi [\pasp] {10.1086/500228}, \href
  {https://ui.adsabs.harvard.edu/abs/2006PASP..118....2H} {118, 2}

\bibitem[\protect\citeauthoryear{Harris et~al.,}{Harris
  et~al.}{2020}]{harris2020array}
Harris C.~R.,  et~al., 2020, \mn@doi [Nature] {10.1038/s41586-020-2649-2}, 585,
  357

\bibitem[\protect\citeauthoryear{{Howell} et~al.,}{{Howell}
  et~al.}{2006}]{2006Natur.443..308H}
{Howell} D.~A.,  et~al., 2006, \mn@doi [\nat] {10.1038/nature05103}, \href
  {https://ui.adsabs.harvard.edu/abs/2006Natur.443..308H} {443, 308}

\bibitem[\protect\citeauthoryear{{Howell} et~al.,}{{Howell}
  et~al.}{2009}]{2009ApJ...691..661H}
{Howell} D.~A.,  et~al., 2009, \mn@doi [\apj] {10.1088/0004-637X/691/1/661},
  \href {https://ui.adsabs.harvard.edu/abs/2009ApJ...691..661H} {691, 661}

\bibitem[\protect\citeauthoryear{{Hsiao}}{{Hsiao}}{2009}]{2009PhDT.......228H}
{Hsiao} Y. C.~E.,  2009, PhD thesis, University of Victoria, Canada

\bibitem[\protect\citeauthoryear{{Hsiao}, {Conley}, {Howell}, {Sullivan},
  {Pritchet}, {Carlberg}, {Nugent}  \& {Phillips}}{{Hsiao}
  et~al.}{2007}]{2007ApJ...663.1187H}
{Hsiao} E.~Y.,  {Conley} A.,  {Howell} D.~A.,  {Sullivan} M.,  {Pritchet}
  C.~J.,  {Carlberg} R.~G.,  {Nugent} P.~E.,   {Phillips} M.~M.,  2007, \mn@doi
  [\apj] {10.1086/518232}, \href
  {https://ui.adsabs.harvard.edu/abs/2007ApJ...663.1187H} {663, 1187}

\bibitem[\protect\citeauthoryear{{Hsiao} et~al.,}{{Hsiao}
  et~al.}{2019}]{2019PASP..131a4002H}
{Hsiao} E.~Y.,  et~al., 2019, \mn@doi [\pasp] {10.1088/1538-3873/aae961}, \href
  {https://ui.adsabs.harvard.edu/abs/2019PASP..131a4002H} {131, 014002}

\bibitem[\protect\citeauthoryear{Hunter}{Hunter}{2007}]{Hunter:2007}
Hunter J.~D.,  2007, \mn@doi [Computing in Science \& Engineering]
  {10.1109/MCSE.2007.55}, 9, 90

\bibitem[\protect\citeauthoryear{{Kasen}}{{Kasen}}{2006}]{2006ApJ...649..939K}
{Kasen} D.,  2006, \mn@doi [\apj] {10.1086/506588}, \href
  {https://ui.adsabs.harvard.edu/abs/2006ApJ...649..939K} {649, 939}

\bibitem[\protect\citeauthoryear{{Kelly}}{{Kelly}}{2007}]{2007ApJ...665.1489K}
{Kelly} B.~C.,  2007, \mn@doi [\apj] {10.1086/519947}, \href
  {https://ui.adsabs.harvard.edu/abs/2007ApJ...665.1489K} {665, 1489}

\bibitem[\protect\citeauthoryear{{Khokhlov}}{{Khokhlov}}{1995}]{1995ApJ...449..695K}
{Khokhlov} A.~M.,  1995, \mn@doi [\apj] {10.1086/176091}, \href
  {https://ui.adsabs.harvard.edu/abs/1995ApJ...449..695K} {449, 695}

\bibitem[\protect\citeauthoryear{{Kim} et~al.,}{{Kim}
  et~al.}{2013}]{2013ApJ...766...84K}
{Kim} A.~G.,  et~al., 2013, \mn@doi [\apj] {10.1088/0004-637X/766/2/84}, \href
  {https://ui.adsabs.harvard.edu/abs/2013ApJ...766...84K} {766, 84}

\bibitem[\protect\citeauthoryear{{Krisciunas} et~al.,}{{Krisciunas}
  et~al.}{2001}]{2001AJ....122.1616K}
{Krisciunas} K.,  et~al., 2001, \mn@doi [\aj] {10.1086/322120}, \href
  {https://ui.adsabs.harvard.edu/abs/2001AJ....122.1616K} {122, 1616}

\bibitem[\protect\citeauthoryear{{Krisciunas}, {Phillips}  \&
  {Suntzeff}}{{Krisciunas} et~al.}{2004}]{2004ApJ...602L..81K}
{Krisciunas} K.,  {Phillips} M.~M.,   {Suntzeff} N.~B.,  2004, \mn@doi [\apjl]
  {10.1086/382731}, \href
  {https://ui.adsabs.harvard.edu/abs/2004ApJ...602L..81K} {602, L81}

\bibitem[\protect\citeauthoryear{{Krisciunas} et~al.,}{{Krisciunas}
  et~al.}{2017}]{2017AJ....154..211K}
{Krisciunas} K.,  et~al., 2017, \mn@doi [\aj] {10.3847/1538-3881/aa8df0}, \href
  {https://ui.adsabs.harvard.edu/abs/2017AJ....154..211K} {154, 211}

\bibitem[\protect\citeauthoryear{{Li} et~al.,}{{Li}
  et~al.}{2003}]{2003PASP..115..453L}
{Li} W.,  et~al., 2003, \mn@doi [\pasp] {10.1086/374200}, \href
  {https://ui.adsabs.harvard.edu/abs/2003PASP..115..453L} {115, 453}

\bibitem[\protect\citeauthoryear{{Maeda}, {Nozawa}, {Nagao}  \&
  {Motohara}}{{Maeda} et~al.}{2015}]{2015MNRAS.452.3281M}
{Maeda} K.,  {Nozawa} T.,  {Nagao} T.,   {Motohara} K.,  2015, \mn@doi [\mnras]
  {10.1093/mnras/stv1498}, \href
  {https://ui.adsabs.harvard.edu/abs/2015MNRAS.452.3281M} {452, 3281}

\bibitem[\protect\citeauthoryear{{Mandel}, {Narayan}  \& {Kirshner}}{{Mandel}
  et~al.}{2011}]{2011ApJ...731..120M}
{Mandel} K.~S.,  {Narayan} G.,   {Kirshner} R.~P.,  2011, \mn@doi [\apj]
  {10.1088/0004-637X/731/2/120}, \href
  {https://ui.adsabs.harvard.edu/abs/2011ApJ...731..120M} {731, 120}

\bibitem[\protect\citeauthoryear{{Mazzali} et~al.,}{{Mazzali}
  et~al.}{2005}]{2005ApJ...623L..37M}
{Mazzali} P.~A.,  et~al., 2005, \mn@doi [\apjl] {10.1086/429874}, \href
  {https://ui.adsabs.harvard.edu/abs/2005ApJ...623L..37M} {623, L37}

\bibitem[\protect\citeauthoryear{{Nobili} et~al.,}{{Nobili}
  et~al.}{2005}]{2005A&A...437..789N}
{Nobili} S.,  et~al., 2005, \mn@doi [\aap] {10.1051/0004-6361:20042463}, \href
  {https://ui.adsabs.harvard.edu/abs/2005A&A...437..789N} {437, 789}

\bibitem[\protect\citeauthoryear{{Nugent}, {Phillips}, {Baron}, {Branch}  \&
  {Hauschildt}}{{Nugent} et~al.}{1995}]{1995ApJ...455L.147N}
{Nugent} P.,  {Phillips} M.,  {Baron} E.,  {Branch} D.,   {Hauschildt} P.,
  1995, \mn@doi [\apjl] {10.1086/309846}, \href
  {https://ui.adsabs.harvard.edu/abs/1995ApJ...455L.147N} {455, L147}

\bibitem[\protect\citeauthoryear{{Perlmutter} et~al.,}{{Perlmutter}
  et~al.}{1999}]{1999ApJ...517..565P}
{Perlmutter} S.,  et~al., 1999, \mn@doi [\apj] {10.1086/307221}, \href
  {https://ui.adsabs.harvard.edu/abs/1999ApJ...517..565P} {517, 565}

\bibitem[\protect\citeauthoryear{{Pessi} et~al.,}{{Pessi}
  et~al.}{2019}]{2019MNRAS.488.4239P}
{Pessi} P.~J.,  et~al., 2019, \mn@doi [\mnras] {10.1093/mnras/stz1855}, \href
  {https://ui.adsabs.harvard.edu/abs/2019MNRAS.488.4239P} {488, 4239}

\bibitem[\protect\citeauthoryear{{Phillips}}{{Phillips}}{1993}]{1993ApJ...413L.105P}
{Phillips} M.~M.,  1993, \mn@doi [\apjl] {10.1086/186970}, \href
  {https://ui.adsabs.harvard.edu/abs/1993ApJ...413L.105P} {413, L105}

\bibitem[\protect\citeauthoryear{{Phillips} et~al.,}{{Phillips}
  et~al.}{2019}]{2019PASP..131a4001P}
{Phillips} M.~M.,  et~al., 2019, \mn@doi [\pasp] {10.1088/1538-3873/aae8bd},
  \href {https://ui.adsabs.harvard.edu/abs/2019PASP..131a4001P} {131, 014001}

\bibitem[\protect\citeauthoryear{{Pskovskii}}{{Pskovskii}}{1977}]{1977SvA....21..675P}
{Pskovskii} I.~P.,  1977, \sovast, \href
  {https://ui.adsabs.harvard.edu/abs/1977SvA....21..675P} {21, 675}

\bibitem[\protect\citeauthoryear{{Riess} et~al.,}{{Riess}
  et~al.}{1998}]{1998AJ....116.1009R}
{Riess} A.~G.,  et~al., 1998, \mn@doi [\aj] {10.1086/300499}, \href
  {https://ui.adsabs.harvard.edu/abs/1998AJ....116.1009R} {116, 1009}

\bibitem[\protect\citeauthoryear{{Stanishev} et~al.,}{{Stanishev}
  et~al.}{2018}]{2018A&A...615A..45S}
{Stanishev} V.,  et~al., 2018, \mn@doi [\aap] {10.1051/0004-6361/201732357},
  \href {https://ui.adsabs.harvard.edu/abs/2018A&A...615A..45S} {615, A45}

\bibitem[\protect\citeauthoryear{{Stritzinger} et~al.,}{{Stritzinger}
  et~al.}{2011}]{2011AJ....142..156S}
{Stritzinger} M.~D.,  et~al., 2011, \mn@doi [\aj]
  {10.1088/0004-6256/142/5/156}, \href
  {https://ui.adsabs.harvard.edu/abs/2011AJ....142..156S} {142, 156}

\bibitem[\protect\citeauthoryear{{Sullivan} et~al.,}{{Sullivan}
  et~al.}{2010}]{2010MNRAS.406..782S}
{Sullivan} M.,  et~al., 2010, \mn@doi [\mnras]
  {10.1111/j.1365-2966.2010.16731.x}, \href
  {https://ui.adsabs.harvard.edu/abs/2010MNRAS.406..782S} {406, 782}

\bibitem[\protect\citeauthoryear{{Tripp}}{{Tripp}}{1998}]{1998A&A...331..815T}
{Tripp} R.,  1998, \aap, \href
  {https://ui.adsabs.harvard.edu/abs/1998A&A...331..815T} {331, 815}

\bibitem[\protect\citeauthoryear{{Uddin} et~al.,}{{Uddin}
  et~al.}{2020}]{2020ApJ...901..143U}
{Uddin} S.~A.,  et~al., 2020, \mn@doi [\apj] {10.3847/1538-4357/abafb7}, \href
  {https://ui.adsabs.harvard.edu/abs/2020ApJ...901..143U} {901, 143}

\bibitem[\protect\citeauthoryear{Virtanen et~al.,}{Virtanen
  et~al.}{2020}]{2020SciPy-NMeth}
Virtanen P.,  et~al., 2020, \mn@doi [Nature Methods]
  {10.1038/s41592-019-0686-2}, \href {https://rdcu.be/b08Wh} {17, 261}

\bibitem[\protect\citeauthoryear{{Wang}, {Li}, {Filippenko}, {Foley}, {Smith}
  \& {Wang}}{{Wang} et~al.}{2008}]{2008ApJ...677.1060W}
{Wang} X.,  {Li} W.,  {Filippenko} A.~V.,  {Foley} R.~J.,  {Smith} N.,   {Wang}
  L.,  2008, \mn@doi [\apj] {10.1086/529070}, \href
  {https://ui.adsabs.harvard.edu/abs/2008ApJ...677.1060W} {677, 1060}

\bibitem[\protect\citeauthoryear{{W}es {M}c{K}inney}{{W}es
  {M}c{K}inney}{2010}]{mckinney-proc-scipy-2010}
{W}es {M}c{K}inney 2010, in {S}t\'efan van~der {W}alt {J}arrod {M}illman eds,
  {P}roceedings of the 9th {P}ython in {S}cience {C}onference. pp 56 -- 61,
  \mn@doi{10.25080/Majora-92bf1922-00a}

\makeatother
\end{thebibliography}








\bsp	
\label{lastpage}
\end{document}